\documentclass[a4paper]{ar-1col}
\usepackage[numbers]{natbib}
\usepackage{amsmath,amssymb,mathrsfs}
\usepackage{bm}
\def\p{{\bm{p}}}
\def\x{{\bm{x}}}

\newcommand{\ks}{{\bm{k}}}
\newcommand{\h}{{\bm{h}}}
\newcommand{\dd}{{\rm d}}
\def\Tform{t_{\rm form}}

\newcommand{\tauhydro}{\tau_\text{hydro}}
\def\st{\begin{equation}}
\def\stp{\end{equation}}
\def\llangle{\left\langle}
\def\rrangle{\right\rangle}

\newcommand{\ps}{{\bm p}}

\newcommand{\qs}{{\bm q}}

\newcommand{\el}{{\scriptscriptstyle \mathrm{el}}}
\newcommand{\inel}{{\scriptscriptstyle \mathrm{inel}}}

\def\alphas{\alpha_s}
\def\LPM{{\scriptscriptstyle \rm LPM}}
\def\BH{{\scriptscriptstyle \rm BH}}
\newcommand{\vs}{{\bm v}}
\newcommand{\ff}{{\bm f}}
\newcommand{\Ps}{{\bm P}}
\def\half{{\tfrac{1}{2}}}
\newcommand{\q}{{\bm q}}
\def\drangle{\rangle\!\rangle}
\def\dlangle{\langle\!\langle}
\def\hard{{\scriptscriptstyle \textrm{hard}}}

\def\psoft{p_{\textrm{soft}}}

\jname{Xxxx. Xxx. Xxx. Xxx.}
\jvol{AA}
\jyear{2019}
\doi{10.1146/((please add article doi))}

\usepackage{bm}
\def\st{\begin{equation}}
\def\stp{\end{equation}}
\def\bg{\begin{eqnarray}}
\def\nd{\end{eqnarray}}
\def\Eq#1{eq.~(\ref{#1})}

\def\eq#1{(\ref{#1})}

\def\Fig#1{Fig.~\ref{#1}}

\def\Sect#1{Sect.~\ref{#1}}
\def\Ref#1{ref.~\cite{#1}}

\def\llangle{\left\langle}
\def\rrangle{\right\rangle}

\usepackage[citecolor=red,linkcolor=red,urlcolor=red]{hyperref}

\begin{document}

\markboth{S.~Schlichting and D.~Teaney}{The First fm/c of Heavy-Ion Collisions}

\title{The First fm/c of Heavy-Ion Collisions}

\author{S.~Schlichting,$^1$ D.~Teaney,$^2$ \affil{$^1$Fakult\"{a}t f\"{u}r Physik, Universit\"{a}t Bielefeld, D-33615 Bielefeld, Germany; email: sschlichting@physik.uni-bielefeld.de }
\affil{$^2$ Department of Physics and Astronomy, Stony Brook University, Stony Brook, New York 11794, USA; email: derek.teaney@stonybrook.edu}}

\begin{abstract}
   We present an introductory review of the early time dynamics of high-energy
   heavy-ion collisions and the kinetics of high temperature QCD. 
   The equilibration mechanisms in the quark-gluon plasma 
   uniquely reflect the non-abelian and ultra-relativistic character of the many body system. 
   Starting with a brief expose of the key theoretical
   and experimental questions, we provide an overview of the theoretical tools
   employed in weak coupling studies of the early time non-equilibrium
   dynamics. We highlight theoretical progress in understanding different
   thermalization mechanisms in weakly coupled non-abelian plasmas, and discuss
   their relevance in describing the approach to local thermal equilibrium
   during the first ${\rm fm}/c$ of a heavy-ion collision. Some important
   connections to the phenomenology of heavy-ion collisions are also briefly
   discussed. 
\end{abstract}

\maketitle

\tableofcontents


\section{Introduction and motivation}

The purpose of ultra-relativistic heavy ion collisions is to produce and 
to characterize the properties of the Quark-Gluon-Plasma (QGP), which is an extreme state of Quantum-Chromo-Dynamic (QCD) matter that was also present in the early universe, during
the first mircoseconds after the big bang.
Over the last two decades, experiments at the Relativistic Heavy-Ion Collider
(RHIC) and  the Large Hadron Collider (LHC) have collided a variety of 
nuclei over a wide range of energies,  and, at least in
the collisions of large nuclei, these experiments show
that  the produced constituents re-interact,
and exhibit  multi-particle correlations with wavelengths which are long 
compared to the microscopic correlation lengths, providing overwhelming
evidence of collective hydrodynamic flow~\cite{Heinz:2013th}.  Hydrodynamic simulations of these 
large nuclear systems describe the observed correlations in exquisite detail with a minimal number of parameters~\cite{Heinz:2013th}.   
In smaller systems such as proton-proton (pp) and proton-nucleus (pA) long range
flow-like correlations amongst the produced particles have also been
observed \cite{Dusling:2015gta,Nagle:2018nvi}, and these observations drive current research into the equilibration
mechanism of the QGP. This research aims to understand how
the observed correlations change with system size, and approach the hydrodynamic 
regime for large nuclei.

Explaining approximately how an equilibrated state of quarks and gluons emerges
from the initial wave functions of the incoming nuclei has been one of the
central goals of the heavy ion theory community for a long time.  Even though
genuinely non-perturbative real-time QCD calculations are currently not
available to address this question (as they suffer from a severe sign problem),
significant progress has been achieved in understanding properties of the
initial state and the equilibration mechanism based on ab-initio calculations
at weak and strong coupling. Here we focus on the weak coupling description, based on the idea that at high energy density and temperatures the coupling constant 
between quarks and gluons $\alphas$ becomes small, and weak coupling methods can be used to analyze the initial production of quarks and
gluons, and the kinetic processes which ultimately lead to a thermalized QGP. When extrapolated to realistic coupling strength, the weak coupling approach based on perturbative QCD and strong-coupling approaches based on the holography yield similar results for the macroscopic evolution of the system~\cite{Keegan:2015avk}. For a recent review of the strong coupling description we refer to \cite{Heller:2016gbp}.

The weakly coupled picture of the equilibration process in
high energy collisions was outlined in a seminal paper by  Baier, Mueller, Schiff and Son (BMSS)~\cite{Baier:2000sb}, and is referred to as the bottom-up thermalization scenario,which is schematically depicted in Fig.~\ref{Overview_Bup}. 
We provide a short review of bottom-up in \Sect{Bottomup}, and then describe recent reanalyses which have clarified and extended the original picture considerably. 
These extensions have turned the parametric estimates of BMSS into hard numbers, which can be used to make contact with the experimental data.

We emphasize that the study of the equilibration mechanisms in non-abelian
gauge theories, such as QCD,  is  of profound theoretical interest, and much of
the research into thermalization is only tangentially driven by the immediate
needs of experimental heavy ion physics program.  In this spirit this review
aims to cover some of the most important theoretical developments regarding the
equilibration mechanism in non-abelian plasmas. Starting with an introductory
discussion of the basic physics picture of the early stages of high-energy
heavy-ion collisions in \Sect{Bottomup}, the subsequent sections,
Sects.~\ref{basics_kinetics} and \ref{basics}, provide a more detailed
theoretical discussion of the underlying theory and the equilibration process
of weakly coupled non-abelian plasmas. New developments based on microscopic
simulations and connections to heavy-ion phenomenology are then discussed in
\Sect{sec:pheno}.

\section{Early time dynamics of heavy-ion collisions}
\label{Bottomup}

When two nuclei collide at high energies, they pass through each other scarcely stopped, leaving behind a debris of highly excited matter which continues to expand longitudinally~\cite{Bjorken:1982qr}. Since
the system is approximately invariant under boosts in the longitudinal ($z$) direction, one point functions of the stress tensor and other fields in the central rapidity region, i.e. the region close to the original interaction point, only depend on proper time $\tau = \sqrt{t^2 - z^2 }$, but do not depend on the space time rapidity $\eta = \tfrac{1}{2} \log((t+z)/(t-z))$.  In co-moving $(\tau,x,y,\eta)$ coordinates, the metric is
\st
  ds^2 = -d\tau^2 + dx^2 + dy^2 + \tau^2 d\eta^2 \, , 
\stp
indicating that the boost-invariant system is continually expanding along the beam axis, $dz= \tau\, d\eta$. While initially the system is in a state far from local thermal equilibrium, phenomenology suggests that on a time $\tauhydro \sim 1 \, {\rm fm/c}$ the plasma of quarks and gluons is sufficiently close to equilibrium  that hydrodynamic constitutive relations are approximately satisfied and the subsequent evolution can be described with hydrodynamics. 

\begin{figure}
   \begin{center}
   \includegraphics[width=\textwidth]{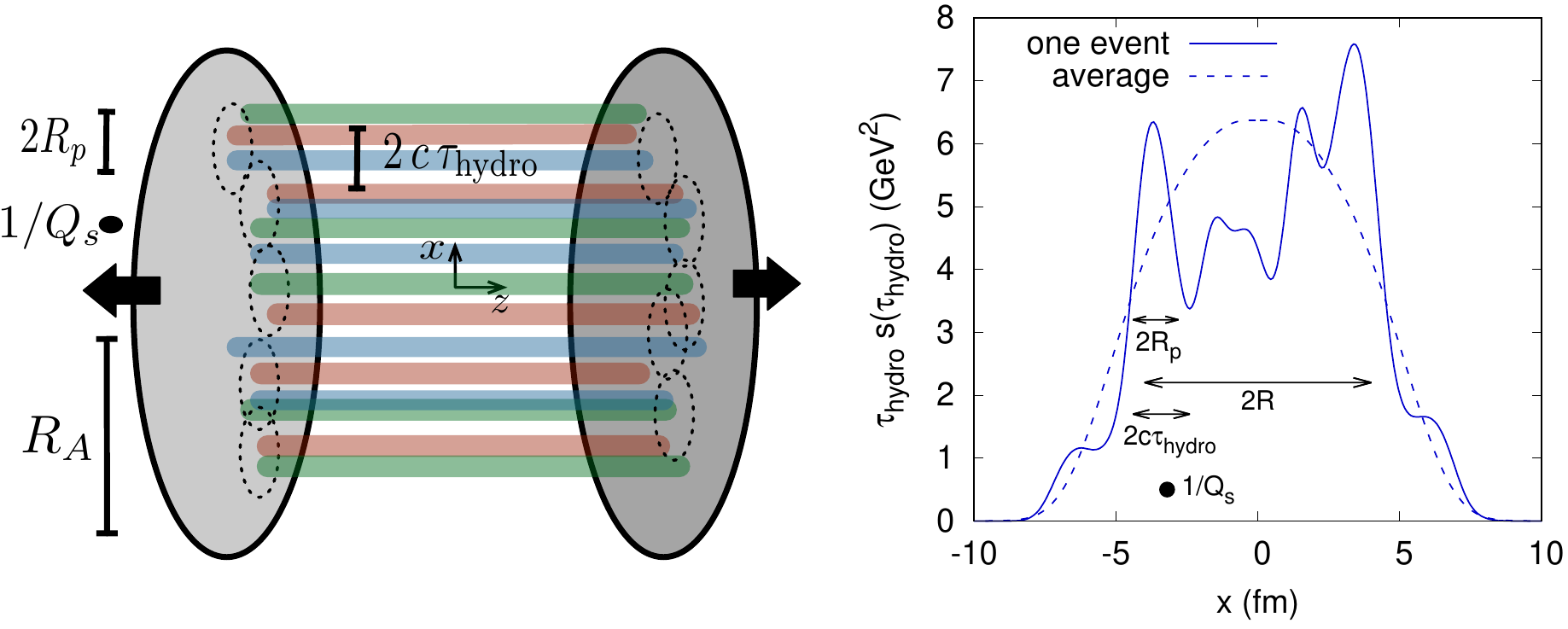}
\end{center}
\caption{(left) Illustration of the two nuclei as they are passing through each
other. Classical color field configurations just after passage were described
in ~\cite{McLerran:1993ni,Kovner:1995ts,Kovner:1995ja}  and feature strong
longitudinal chromo-electric and chromo-magnetic fields, which rapidly decohere
on a timescale of $\sim Q_s$~\cite{Lappi:2006fp}. (right) Snapshot of a typical
entropy density profile used in event-by-event hydrodynamic simulations of
heavy-ion collisions~\cite{Mazeliauskas:2015vea}. Smaller scale fluctuations on microscopic length
scales $\sim1/Q_s$ are not shown, and are indicated by the black
dot. }
\label{geometry}
\end{figure}

While the longitudinal structure is approximately homogenous in space time
rapidity, the transverse structure of the fireball is always inhomogenous,
reflecting the initial geometry of the collision. In \Fig{geometry} we show
a typical transverse (entropy density)
profile that is used to initialize  hydrodynamic simulations of the space time
evolution. While the average geometry is characterized by the nuclear radius
$R_A$, one finds that in any realistic event-by-event simulation
there are smaller length scales  in the initial geometry of order the proton
radius, $R_p \ll R_A$,  which arise from fluctuations in the
positions of the incoming protons. Such geometric fluctuations are responsible
for many of the most prominent flow observables in heavy ion collisions such as
e.g. the triangular flow~\cite{Gelis:2016upa}. Still smaller fluctuations of
order the inverse saturation momentum $Q_s^{-1}$ (see \Sect{sec:introsat}) are not shown in
this figure. Different scales in \Fig{geometry} should be compared to the
distance scale $c \tauhydro$, which provides an estimate of the causal
propagation distance during the approach to equilibrium. We will generally
assume that $c\tauhydro$ is short compared to the nuclear radius, $c\tauhydro
\ll R_A$, such that on average the transition from the non-equilibrium state
towards thermal equilibrium proceeds locally in space and can discussed at the
level of individual cells of size $c\tauhydro$. Short distance fluctuations on
scales $c\tauhydro$ spoil this picture; however such effects were neglected in
the original bottom-up scenario and we will follow this assumption by
approximating the evolution of the system as homogenous in transverse space and
space time rapidity throughout most of this review. Shortcomings of this
approximation will be discussed further in Sec.~\ref{sec:pheno} and
\ref{sec:epilogue} along with recent extensions of the original work of BMSS, which
incorporate short distance fluctuations of the nucleon positions on scales
$R_{p} \sim c\tauhydro$ into the description of the first fm/c of heavy-ion
collisions.

\subsection{Microscopics of the initial state}
\label{sec:introsat}
In each small circle of size $c \tauhydro$ in the transverse plane the initial production of quarks and gluons in 
momentum space follows from the Color-Glass-Condensate (CGC) effective theory of parton
saturation~\cite{Iancu:2003xm,Gelis:2010nm}. Briefly, in this theory the
incoming nuclei are highly length contracted by an ultra-relativistic factor $\gamma \gg 1$, and the density of
gluons per transverse area and rapidity in the wave functions of the nuclei, 
$(dN/dy)/\pi R_A^2 $,
grows with increasing collision energy. Here 
$dN/dy$ is the number of gluons per rapidity $y$ which is related to Bjorken $x_{\rm bj}$, $dy = dx_{\rm bj}/x_{\rm bj}$. 
This transverse density of gluons 
determines a momentum scale,  known as the saturation momentum $Q_s$, 
which at very high energies can become large compared to $\Lambda_{\rm QCD}$
\st
Q_s^2   \sim  \frac{\alphas}{\pi R_A^2} \frac{dN}{dy } \gg \Lambda_{QCD}^2 \, .
\stp
The saturation momentum $Q_s$ sets the momentum 
scale for the transverse momentum distribution of partons in the  wave functions. 
For $Q_s \gg \Lambda_{QCD}$ the coupling constant is small $\alphas(Q_s) \ll 1$,  
and the  evolution of the system can be treated using weakly coupled methods.  
Further, the number of gluons per phase space cell in the incoming wave functions is large
\st
\frac{1}{\pi Q_s^2 R_A^2}  \frac{dN}{dy} \sim \frac{1}{\alphas}  \gg 1 \,  ,
\stp
and in this regime the evolution of the system is classical.  
Thus, the production of gluons and their initial evolution system is determined by solving the non-linear classical Yang-Mills equations of motion~\cite{McLerran:1993ni,Kovner:1995ts,Kovner:1995ja}.  In practice, the saturation momentum is  $Q_s\sim 1\,{\rm GeV}$ at RHIC and $2\,{\rm GeV}$ at the LHC. As these values not vastly larger than $\Lambda_{QCD}$ there will always
be important quantum corrections to the CGC formalism, which will almost be completely neglected in this review.

In an important set of papers,
the initial conditions for the classical fields in the forward light
cone just after the 
intitial crossing of the two nuclei were worked out (analytically) by
matching the classical fields just  before the collision 
with those just after crossing~\cite{McLerran:1993ni,Kovner:1995ts,Kovner:1995ja}. 
These initial conditions consist of strong longitudinal fields, $E^z$ and $B^z$, which as illustrated in \Fig{geometry} is somewhat reminiscent of a parallel plate capacitor~\cite{Lappi:2006fp}. Indeed,
the average stress tensor for a boost invariant, or Bjorken, expansion  and a conformal system 
(with $T^{\mu}_{\; \mu}=0$) must take the form
$ \llangle T^{\mu}_{\;\nu} \rrangle = (-\epsilon,P_T, P_T, P_L),
$ with $\epsilon=2  P_T + P_L $.  
The  matching procedure~\cite{McLerran:1993ni,Kovner:1995ts,Kovner:1995ja} shows that  $\llangle T^{\mu}_{\;\nu} \rrangle =
(-\epsilon,\epsilon,  \epsilon,  -\epsilon )$, and thus, the initial
longitudinal ``pressure''  $P_L$ is negative as  for a
constant electric (or magnetic) field in the $z$-direction in classical electrodynamics. 
These strong longitudinal fields rapidly decrease on a time scale of $\sim Q_s$ as the classical field configuration decoheres. 

The initial conditions outlined in the preceding paragraph motivated the first classical simulations of gluon production in the longitudinally expanding  boost invariant geometry~\cite{Krasnitz:1999wc,Krasnitz:2000gz}.  
In the original formulations the classical fields were assumed to remain effectively 2+1 dimensional, i.e. strictly independent of rapidity as a function of 
time $\tau$, reflecting the fact that the initial conditions are boost invariant up to quantum corrections of order $\alphas$.
However, such quantum fluctuations provide the seed from which classical instabilities develop in the longitudinal direction~\cite{Epelbaum:2013waa,Gelis:2016upa}, such that the gluonic fields quickly become chaotic in all three
dimensions and the classical solutions are only rapidity-independent on average.  The instabilities grow as $~e^{\Gamma\sqrt{Q_s \tau}}$,  with $\Gamma \sim 1$, limiting the applicability of strictly boost invariant simulations to short times,  $\tau  \lesssim Q_s^{-1} \, \log^2 (1/\alphas)$~\cite{Romatschke:2005pm,Romatschke:2006nk,Berges:2012cj}.
In spite of this shortcoming, strictly boost invariant simulations of classical field dynamics form the basis of phenomenological studies of particle production and early time dynamics in the IP-Glasma model ~\cite{Schenke:2012wb,Schenke:2012fw}.  

During the classical evolution the field strength decreases due
to the longitudinal expansion, and eventually the equations of motion linearize.
For times long enough  $\tau Q_s \gg 1$
(but not too long; see \Sect{sec:bupintro}) the phase space density of gluons is still large but much smaller than the inverse self-coupling $\alphas^{-1}(Q_s)$.
In this regime, either kinetic theory or classical field theory
can be used to simulate the evolution of the system~\cite{Mueller:2002gd,Aarts:1997kp,Jeon:2004dh}. 
In particular, it is sensible to talk about the gluon phase space distribution, as opposed to the classical field configuration. 
The initial phase space distribution of gluons
$f(\tau,\x,\p)$ can be determined
from the classical simulations by evaluating the Wigner transform of equal time two point functions of gauge fields, after fixing a physical gauge such as the Coulomb Gauge (see for instance \Ref{Berges:2012ev,Berges:2013eia,Greif:2017bnr}).   
Due to the longitudinal expansion of the system, the initial phase-space distribution of the system is strongly squeezed with $\llangle p_\perp^2 \rrangle \sim Q_s^2$ and
$\llangle (p^z)^2 \rrangle   \ll  \llangle p_\perp^2 \rrangle$.

\subsection{Bottom-up equilibration}
\label{sec:bupintro}

\begin{figure}
     \includegraphics[width=\textwidth]{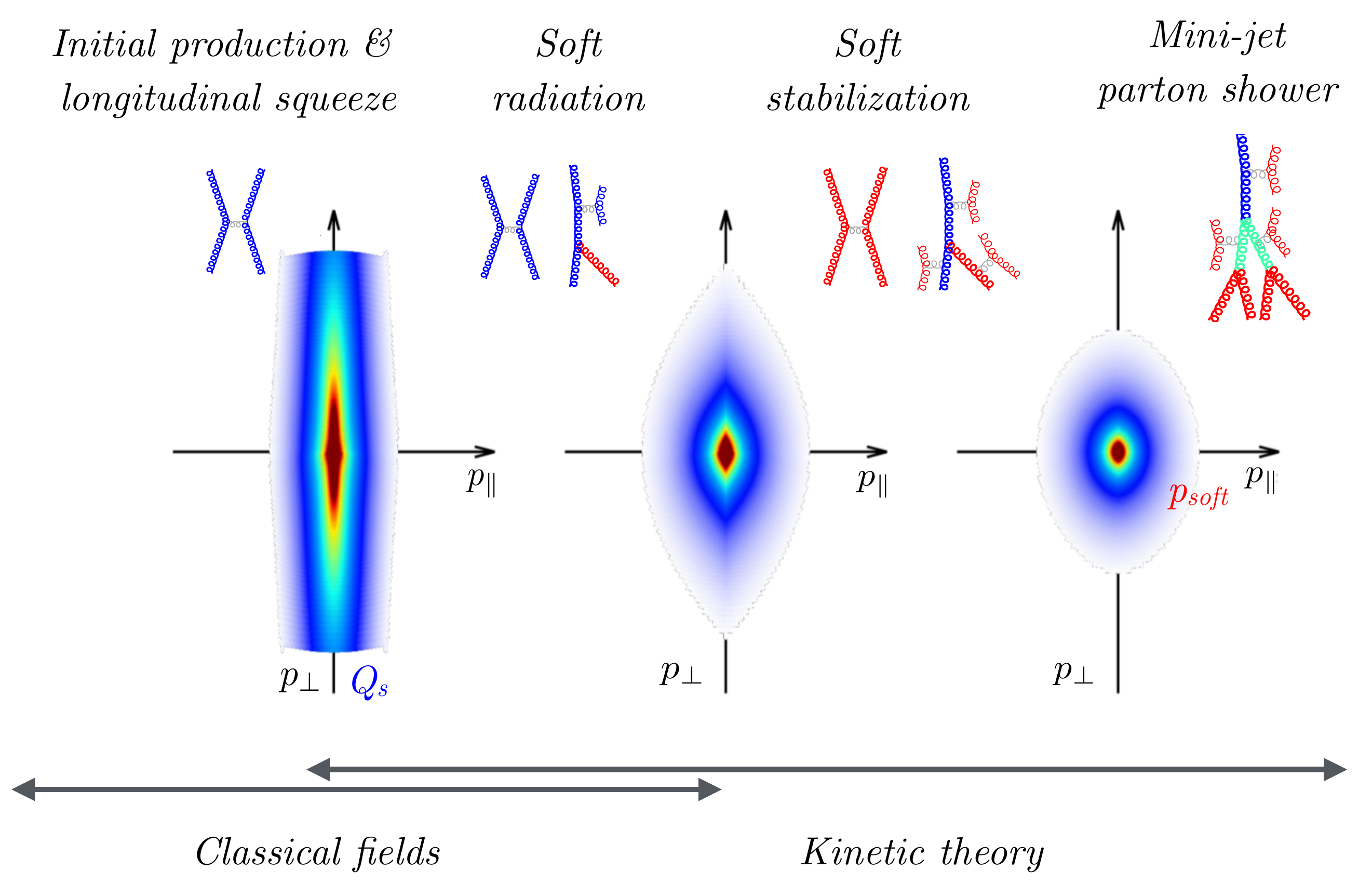}
      \caption{\label{Overview_Bup}  Schematic overview of the bottom-up thermalization showing the evolution of the phase-space distribution of gluons in momentum space based on kinetic theory simulations of ~\cite{Kurkela:2018vqr}. Different regimes correspond to evolution times $\tau/\tauhydro\approx0.1,~0.5,~1$ at realistic coupling strength $\alphas \approx 0.3$ }
\end{figure}

This highly anisotropic initial state provides the starting point for the bottom-up scenario, which is illustrated in \Fig{Overview_Bup}.   During the first classical phase of bottom up the phase space distribution becomes increasingly anisotropic as
time progresses.

In the original bottom-up proposal,
the longitudinal width of the phase space distribution $\llangle p_z^2 \rrangle$ is 
determined by momentum diffusion, i.e. small angle scatterings amongst the hard particles.  
The diffusion process tries to increase the longitudinal width,
but competes with the expansion of the system. This competition leads to a
scaling solution for the phase space distribution $f(\tau, p_z, p_\perp)$ at
late times $Q_s \tau \gg 1$,
where the  transverse and longitudinal momenta are of order
\begin{subequations}
\label{scaling_phase1}
\begin{align}
   \llangle p_T^2 \rrangle \sim& Q_s^2\, ,  \\
   \llangle p_z^2 \rrangle \sim& \frac{Q_s^2 }{(Q_s\tau)^{2/3} } \, .
\end{align}
\end{subequations}
During the first stage of bottom-up, the number of hard gluons per rapidity remains constant $dN/dy \sim Q_s^2 R_A^2/\alphas$, and thus the density of
gluons (the number per volume) decreases as $n_h \sim Q_s^2/\alphas \tau$ due to the expansion of the system.  Based on these
estimates, the phase space density of hard modes 
decreases as
\st
\label{bupphase1f}
f_h \sim  \frac{1}{\alphas}  \frac{1}{(Q_s \tau)^{2/3} } \,  ,
\stp
following a pattern which is characteristic of overoccupied initial states with $f_h \gg 1$, 
which will be  discussed in greater detail in \Sect{basics_overoccupied} and \Sect{bupwnumbers}.
Analyzing \Eq{bupphase1f}, we see that  the phase space density becomes of
order unity at a time of order $Q_s \tau \sim \alphas^{-3/2}$, marking the end of the first over-occupied stage. 
Most importantly, from this point onward the system can no longer be treated as a classical field,
and its subsequent evolution must be analyzed with kinetic theory.

In the second stage of bottom-up, $Q_s\tau \gg \alphas^{3/2}$, 
radiation from the hard modes increases
the number of soft gluons per rapidity.  
Ultimately this soft bath will thermalize the hard modes 
giving the bottom-up equilibration scenario its name.
While the soft bath is being populated, the number of hard particles per volume continues to decrease due to the longitudinal expansion,  $n_h \sim Q_s^2/\alphas \tau$. Now, however, the longitudinal width  $\llangle p_z^2 \rrangle$
of these hard modes remains constant in 
time, since the increase in width from (momentum) diffusion is compensated for by the expansion of the system
\begin{subequations}
   \label{bup-phase2}
\begin{align}
   \llangle p_T^2 \rrangle  \sim &  Q_s^2 \, ,  \\
   \llangle p_z^2 \rrangle  \sim &  \alphas Q_s^2 \, .
\end{align}
\end{subequations}
Thus, the phase space density of hard particles in the second phase decreases  as 
\st
\label{bup2ndphase}
f_h \sim   \frac{1}{\alphas^{3/2}} \frac{1}{(Q_s\tau) }  \, , \qquad Q_s \tau \gg \alphas^{-3/2}   \, ,
\stp
and is therefore  much smaller than unity.  Indeed,
at the end of the second phase of bottom-up, $Q_s\tau \sim \alphas^{-5/2}$,
the phase space density of the hard modes is parametrically small, $f_{h} \sim \alphas \ll 1$.

In the last stage of the bottom-up $Q_s\tau \gg \alphas^{-5/2}$ the soft bath has equilibrated,
and begins to influence the evolution of the hard particles.  
In this stage there is a cascade of energy  from the scale of $Q_s$ to 
the soft scale scale set by the temperature of the bath.  The physics of this process is analogous to the stopping of ``jets'' with momentum of order $Q_s$ in plasma~\cite{Baier:2000sb,Arnold:2009ik,Blaizot:2013hx} and described further in \Sect{basics_underoccupied} and \Sect{bupwnumbers}. 

The second and third  stages of
the bottom-up scenario are characteristic of the thermalization of initially 
\emph{under-occupied} systems.  
We will see in \Sect{basics_underoccupied} that the buildup of a soft thermal bath, and cascade of energy to the infrared are to be expected in
such systems.

\section{QCD Kinetics: a brief review}
\label{basics_kinetics}

Having qualitatively described the bottom-up picture, we will now turn to
a more quantitative analysis of the equilibration process of the QGP in the framework of kinetic theory. Kinetic processes in the  QGP are markedly different from other many-body systems of condensed matter physics, uniquely reflecting the
non-abelian and ultra-relativistic character of the produced quark and gluon
quasi-particles. 
A complete leading order description of QCD kinetics (close to equilibrium) was given in \cite{Arnold:2002zm}, and
was then used to compute the transport coefficients of the QCD plasma to
leading order in the strong coupling constant~\cite{Arnold:2003zc}. 


Here we will provide a brief review of QCD kinetics to
establish notation  and to collect the principal results.
If not stated otherwise we will focus on pure gauge systems, and
refer to the literature for additional details~\cite{Arnold:2002zm,Ghiglieri:2018dib}.
Further we will, at points, have to assume that the momentum distribution
is isotropic; issues which arise in the description of anisotropic systems (such as plasma instabilities) will be discussed briefly in \Sect{bupwnumbers}. 

The QCD Boltzmann equation takes the form 
\st
\label{eq:QCDBoltzmann}
\left(\partial_t + \vs_\p \cdot \partial_\x \right) f(t,\x,\p)
= C^{2\leftrightarrow 2}[f(\p)] + C^{1\leftrightarrow2}_{\rm inel}[f(\p)] \, ,
\stp
where the $2\leftrightarrow 2$ rates describes elastic scattering, and $C^{1\leftrightarrow 2}$ describes collinear radiation.  We further introduce two dimensionful integrals\footnote{ 
We follow standard notation, where  $d_A = N_c^2 -1$ is the dimension
of the adjoint, while $C_A = N_c$ is its Casimir. By $\nu_g = 2 d_A$ we denote
the number of gluonic degrees of freedom. Phase space integrals are abbreviated as $\int_p \equiv \int d^3p/(2\pi)^3$, as is the phase space density $f_\p \equiv f(t,\x,\p)$.
}
\begin{align}
   \label{eq:scalesmstar}
   m^2 \equiv& \nu_g  \frac{g^2 C_A }{d_A } \int_\p \frac{f_\p}{p} \, , \\
   2 T_* m^2 \equiv& \nu_g  \frac{g^2 C_A }{d_A } \int_\p f_\p (1 + f_\p) \, .
\end{align}
to characterize the momentum distribution. Modes of order $m$ are considered soft, while modes of order $T_*$ are hard. 
In equilibrium $T_*$ is the temperature of the medium, and $m$ is the asymptotic mass of the gluon dispersion curve, i.e.
$E_\p = \sqrt{\p^2 + m^2} \simeq |\p| + m^2/2|\p|$.

\subsection{Elastic scattering and momentum diffusion}

The $2\leftrightarrow 2$ processes can be divided 
into soft collisions,  where the momentum transfer is of order $m$ and screening is important, and hard collisions, where the momentum transfer is above a cutoff scale $\mu^2_\perp \sim T_*m$ and screening can be neglected:
\st
\label{eq:C22all}
C^{2 \leftrightarrow 2}[f(\p)] = C_{\rm diff}[f(\p)] + C_{\rm hard}^{2\leftrightarrow 2} [f(\p)] \, .
\stp
Hard collisions (which are conceptually straightforward) exhibit the same parametric dependencies as soft interactions (see e.g. \cite{Kurkela:2011ti}) and will be ignored in the estimates below. 
Elastic interactions with soft momentum transfers create drag and diffusion processes in momentum space,  which may be summarized by a Fokker-Planck equation. 
This separation into hard and soft collisions was essential  to an almost complete next-to-leading-order computation of the shear viscosity~\cite{Ghiglieri:2018dib}.

Consider a particle of momentum $\p$ (with four velocity $v_\p^{\mu} \equiv (1, \hat \p)$) being jostled by a soft random external field $A^{\mu}(Q)$ created by all other particles.  The absorption rate of three momentum $\q$ by the field is
\st
\label{absorptionrate}
  \frac{ d\Gamma_{\rm el}^>(\hat \p)}{d^3q }  
 = g^2 C_A  \, \int \frac{dq^0}{2\pi} \, v_\p^{\mu} v_\p^{\nu} \; \dlangle A_{\mu}(Q) \, (A_{\nu}(Q))^* \drangle^>  \; 2\pi \delta (v_\p \cdot Q) \, ,
\stp
where the $\delta$-function stems from energy conservation, $2\pi \delta(q^0 + E_{\p-\q}  - E_\p) \simeq 2\pi \delta(v_\p \cdot Q)$. Statistical fluctuations of the gauge field fluctuations are given by 
\st
\label{Wightmannflucts}
\dlangle A^{\mu}(Q) (A^{\nu}(Q))^* \drangle^>  =    G_{R}^{\mu\alpha}(Q) \, \Pi^>_{\alpha \beta}(Q) \, (G_{R}^{\beta \nu}(Q))^* \, ,
\stp
where the Wightman self energy reads
\begin{align}
   \label{piequation}
   (\Pi^>(Q))^{\alpha \beta} =& 
   \nu_g \frac{g^2 C_A}{d_A}
   \int_{\ks} \, v_{\ks}^{\alpha} v_{\ks}^{\beta} 
   \,  f(\ks) (1 + f(\ks+\q) ) \, 
   2\pi 
   \delta(v_\ks \cdot Q) 
   \, ,
\end{align}
Here $G_R (Q) \sim 1/Q^2$ is the hard thermal loop retarded response function~\cite{Blaizot:2001nr},
which can only be worked out in closed form for isotropic systems. 
In the limit of small $\q$ the population factors in \Eq{piequation} become $f(\ks) (1 + f(\ks))$,
and the correlator in \Eq{Wightmannflucts} has a simple interpretation --   it is the correlation amongst the 
gauge fields $A =  (G_{R}(Q)) \cdot \, g v_{\ks}$ produced by random fluctuations
of the phase space density $\delta f(t,\x, \ks)$,
which have the usual equal time Bose-Einstein statistics~\cite{LandauStatPart1}
\st
\dlangle \delta f(t,\x,\ks) \, \delta  f(t,\x',\ks) \drangle =  f(t,\x,\ks)\,  (1 + f(t,\x, \ks) )\, \delta^3(\x - \x') \, (2\pi)^3 \delta^3(\ks - \ks') \, .
\stp



The absorption rate in \Eq{absorptionrate} gives the rate that momentum $\q$ 
is taken from the particle and given to the bath. Similarly, the emission rate
takes the same form  as \Eq{absorptionrate} but replaces the self energy $\Pi^>$ with 
\st
(\Pi^<(Q))^{\alpha \beta} = \nu_g \frac{g^2 C_A}{d_A} 
\int_\ks v_\ks^{\alpha} v_\ks^{\beta} \,  f(\ks+\q) (1 + f(\ks) ) \, 2\pi \delta(v_\ks \cdot Q) \, ,
\stp
such that at small  $\q$  
the emission and absorption rates  are equal, and it is the symmetric correlator $\Gamma_{\rm \el} {=} (\Gamma^>_{\rm \el}+\Gamma_{\rm \el}^<)/2$ that will determine the rates of momentum diffusion below.
Conversely, the difference in the emission and absorption rates determines the
drag, and involves:
\begin{align}
   (\Pi^>(Q) - \Pi^<(Q))^{\alpha\beta}  =&  \nu_g \frac{g^2 C_A}{d_A} \int_\ks \,  v_\ks^{\alpha}  v_\ks^{\beta} \;  q^i\frac{\partial f(\ks)  }{\partial k^i} \,  2\pi \delta (v_\ks \cdot Q)  \,  , \\ 
   = & 2 q^0 \, m^2 \int \frac{d\Omega}{4\pi} \,  v_{\ks}^{\alpha} v_\ks^{\beta} \,     2\pi \delta(v_\ks \cdot Q) \, ,
\end{align}
where in passing to the last line we have assumed that the system is isotropic, $\partial f/\partial k^i = f'(k) \hat k^i$,
allowing us to perform an integration by parts. 



The evolution of the system due to 
soft scattering is a competition between the emission and absorption rates
\st
 \partial_t f_\p   + \vs_\p \cdot \partial_\x f_\p = \int \dd^3q\,    
 \left( \frac{\dd\Gamma_{\rm el}^<(\hat \p)}{\dd^3q} \, f_{\p-\q}  (1 + f_\p)  
 -  \frac{\dd\Gamma_{\rm el}^>(\hat \p)}{\dd^3q} \, f_\p (1 + f_{\p - \q}) \right) \, .
\stp
We will now generally assume that the distribution is isotropic  which
simplifies the analysis of momentum diffusion.
Expanding in powers of the momentum transfer $\q$ (which is small compared to
the momentum $\p$ of the hard particle), we see that the contribution of small angle elastic
processes to the Boltzmann equation (\ref{eq:QCDBoltzmann}) takes the form of a
Fokker-Planck equation
\st
\label{eq:FPequation}
C_{\rm diff}[f(\p)] = \eta^i(\hat \p) 
   \frac{\partial}{\partial p^i} \left( f_\p (1 + f_\p)\right)  +  \hat q^{ij}(\hat \p)  \frac{\partial^2 f_\p }{\partial p^i \partial p^j }  \, ,
\stp 
where the drag and diffusion coefficients are given by
\begin{align}
   \eta^i=& \int \dd^3q  \left(\frac{\dd\Gamma^>_{\rm el}(\hat\p)}{\dd^3q}  - \frac{\dd\Gamma^<_{\rm el}(\hat\p)}{\dd^3q}    \right) q^i \, ,  \\
\hat q^{ij}(\hat\p)  =& \int d^3q  \,  \left(\frac{\dd \Gamma_{\rm el}(\hat \p)}{\dd^3q} \right) q^i q^j  \, .
\end{align}
%
Specifically for isotropic systems these coefficients can be decomposed
as
\st
\eta^i(\hat \p) = \eta \hat p^i \, , \qquad   
q^{ij}(\hat \p) =  \hat q_{L} \hat p^i \hat p^j  + \tfrac{1}{2} \hat q \left( \delta^{ij} - \hat p^i \hat p^j \right)  \, ,
\stp
and the scalar coefficients $\eta,\hat{q}_{L},\hat{q}$ can be evaluated as~(see \cite{Ghiglieri:2015zma} for a review),
\begin{subequations}
   \label{FPcoefficients}
\begin{align}
   \label{eq:etafp}
   \eta =& \frac{g^2 C_A m^2 }{8\pi }  \log \left(\frac{\mu_\perp^2}{m^2} \right) \, ,  \\
   \label{eq:qlfp}
\hat q_L =& \frac{g^2 C_A \, (2 T_* m^2) }{8\pi }  \log\left( \frac{\mu_\perp^2 }{m^2}  \right) \, , \\
   \label{eq:qperpfp}
   \hat q =& \frac{g^2 C_A \, (2 T_* m^2) }{4\pi }  \log\left( \frac{\mu_\perp^2 }{2m^2 }\right) \, .
\end{align}
\end{subequations}
Similarly, the elastic scattering rate for kicks transverse to the direction of the particle can also be evaluated in closed form yielding
\st
\label{eq:Gammael}
(2\pi)^2 \frac{d\Gamma_{\rm el} }{d^2 q_\perp} = g^2 C_A T^{*} \left( \frac{1}{q_\perp^2} - \frac{1}{ q_\perp^2 + 2 m^2} \right) \, .
\stp
Although the Fokker-Planck coefficients in \Eq{eq:etafp} depend on the cutoff scale $\mu_\perp$, 
the time the evolution of the system is independent of $\mu_\perp$,
when both the hard collisions and the Fokker-Planck evolution
are taken into account~\cite{Ghiglieri:2015ala}. We finally note that from
\Eq{eq:Gammael} and \Eq{eq:qperpfp}, the elastic scattering rate is of order

 \st
 \Gamma_{\rm el} \sim \int_{\sim m} \dd^2q_\perp \frac{\dd\Gamma_{\rm el}}{\dd^2q_\perp}  \sim  \frac{\hat q}{m^2} \,,
 \stp
which will be used repeatedly when estimating the rate of collinear radiation described
in the next section.


\subsection{Collinear radiation}
\label{sec:collinear_rad}

Elastic scatterings of ultra-relativistic particles induce collinear radiation as the charged
particles are accelerated by the random kicks from the plasma.
A massless gluon  with momentum $\Ps=\p + \ks$ can split into  
two particles with momentum fractions $z$ and $\bar{z}\equiv (1-z)$, where
$\p=z \Ps$ and $\ks=\bar{z} \Ps$ respectively. 
These radiative process should be incorporated into the Boltzmann 
equation at leading order~\cite{Baier:2000sb,Arnold:2002zm}.   Denoting the rate for this process as
$\dd\Gamma_{\rm inel}(\Ps)/\dd z$, the contribution to the Boltzmann equation can be 
written as\footnote{
Our notation for inelastic splitting rate follows \cite{Arnold:2008iy,Arnold:2008vd}.
Arnold, Moore, and Yaffe use a different symbol $\gamma^{g}_{gg}(\p',\p, \ks)$~\cite{Arnold:2002zm},
which is related to  the rate used here through
\st
\frac{d\Gamma_\inel (\Ps)}{dz} = \frac{(2\pi)^3}{\nu_g |\Ps|} \gamma_{gg}^g(\Ps,z\Ps,(1-z) \Ps) \, .
\stp
}
\begin{align}
\label{C12}
C^{1 \leftrightarrow 2}[f(\p)]
    =& \nu_g \int_{\Ps} \int_0^1 dz  \, \frac{\dd \Gamma_{\inel}(\Ps)}{\dd z} \,%
    \frac{(2\pi)^3 }{\nu_g} \delta^{(3)}(\p - z\Ps)  
    \, \nonumber \\
     & \qquad  \times\left[  f(\Ps) (1 + f(z\p)) (1 +
    f(\bar{z}\Ps))   - f(z\Ps) f(\bar z \Ps) (1 + f(\Ps)) \right]  \nonumber \\
  -&  \frac{1}{2}    \int_0^1 dz  \frac{ d\Gamma_{\inel}(\p) }{dz}  \nonumber \\
   &  \qquad \times  \left[f(\p) (1 + f(z \p)) (1 + f(\bar{z} \p)) - f(z\p) f(\bar{z} \p) (1 + f(\p)) \right] \,,
\end{align}
and we will now briefly describe the characteristic features of the splitting rate. 

In the splitting process the energy difference between the incoming and outgoing states is
\begin{align}
\label{deltaE}
   \delta E =&  E_\p + E_\ks   - E_{\p + \ks}
  \simeq \frac{h^2}{2 P z (1 - z) }   + \frac{m^2}{2 P z}  + \frac{m^2}{2 P ( 1 - z) }  -   \frac{m^2 }{2 P } \, ,
\end{align}
where $\h \equiv z \ks_\perp - (1-z) \p_\perp$ is essentially the transverse momentum 
of the softest fragment. In writing \Eq{deltaE} we have expanded  
the quasiparticle energy for small transverse momentum, $E_\p \simeq p^z + (m^2 + p_\perp^2)/2p$.
Since the Hamiltonian time evolution of the system involves phases of the form $e^{-i \delta E  t}$, 
the splitting process is only completed on a time scale
\st
\label{tformest}
\Tform  \equiv \frac{1}{\delta E} \, .
\stp
which defines an important timescale for collinear radiation, namely the \emph{formation time}. 

For highly energetic particles  the formation time can become long  compared to
the time between elastic collisions.  In this regime multiple scattering 
will suppress the emission of gluon radition,  and this suppression is known as the Landau-Pomenanchuk-Migdal (LPM) effect.   

Let us estimate the energy  $\omega_{\LPM}$   when  
the LPM effect becomes operative, i.e. when $\Tform  \Gamma_{\rm el}  \sim 1$. 
To this end,  consider a splitting process with $z\ll 1$
, so that $h\simeq p_\perp$  
and $p=z P \sim
\omega_{\LPM}$.  In this regime the formation time is of order
\st
\Tform \sim \frac{2p}{p_\perp^2} \sim  \frac{\omega_\LPM}{m^2} \, ,   
\stp
were we have estimated $p_\perp^2 \sim m^2$ as the typical momentum 
associated with a \emph{single} elastic scattering event.
Since the elastic scattering rate is of order $ \Gamma_{\rm el} \sim \hat q/m^2$, we find
\st
\omega_{\LPM } \sim  \frac{m^4 }{\hat q } \, .
\stp
For high energy particles the formation time becomes much longer than $\Gamma_{\rm el}^{-1}$.  In this limit the accumulated transverse momentumg grows as $h^2 \sim  \hat q  \, t_{\rm form} \gg m^2$,  and thus using \Eq{tformest} and \Eq{deltaE} 
we find the following estimate for the formation time
\st
    \Tform \sim \sqrt{ \frac{P} {z (1- z) \hat q } } \, .
\stp

For  $\omega \gtrsim  \omega_{\LPM}$ the radiation rate must account for the multiple scatterings that happen  during the formation time of the radiation.
Conversely, in the Bethe-Heitler (BH) limit $\omega \ll \omega_{\LPM}$, the interference between the scattering events can be neglected, and 
 each scattering   has a probability of order $\alpha$ to radiate  a gluon with momentum
 fraction $z$ disributed according to the  splitting function\footnote{Generally the splitting function for $g\leftrightarrow gg$ is given by $P_{g \rightarrow g}(z) = C_A \frac{1 + z^4  +(1-z)^4}{z (1 - z)}$. However we will frequently approximate  $P_{g \rightarrow g}(z)$ by its soft limit $P_{g \rightarrow g}^{\rm soft}(z)=\frac{2C_A}{z(1-z)}$.} $P_{g\rightarrow g}(z)$. Since the scattering rate is  $\Gamma_{\rm el} \sim \hat q/m^2$, the total splitting rate  in the BH limit is of order
\st
\label{bhestimate}
\frac{\dd \Gamma_{\rm inel}^{\BH} (\p_0)}{\dd z}  \sim \alpha \, P_{g \rightarrow g}^{\rm soft} (z) \,  \frac{\hat q}{m^2 }  \, .
\stp 
More generally emissions radiated within a formation time will 
destructively interfere,  and the net emission rate is determined
by solving an integral equation.
This rate takes the form~\cite{Baier:1996kr,Zakharov:1997uu,Baier:2000sb,Arnold:2002zm,Arnold:2008iy}
\st
\label{eq:splittingrate} 
 \frac{d\Gamma_{\rm \inel}(\Ps)}{\dd z} =   \alphas P_{g \rightarrow g}(z)  
   \int \frac{d^2h}{(2\pi)^2 }  \, \frac { 2 \h \cdot  \text{Re} \ff(\h)    } {(2 P z (1 - z))^2 }   \, ,
\stp
where the integral in this equation has units $({\rm time})^{-1}$. 
The function $\ff(\h)$ (which encodes the current-current statistical correlation function)  satisfies an integral equation of the form
\begin{align}
\label{eq:integraleq} 
 2\h =& i \, \delta E(h) \, \ff(\h) +  \int d^2q_\perp \frac{d \Gamma_{\el}}{d^2q_\perp}
 \big\{\half\left[\ff(\h) -  \ff(\h + \q_\perp) \right] \nonumber \\
  & \qquad \qquad  
  + \half\left[\ff(\h) - \ff(\h + z \q_\perp) \right] + \half \left[ \ff(\h) - \ff(\h + (1- z)\q_\perp) \right] \big\} \, .
\end{align}

To analyze  this equation, let us take the Bethe-Heitler limit when the radiation is 
soft, $ z\ll 1$ and  $\omega \ll \omega_{\LPM}$, so that the formation time is small compared to the elastic scattering rate, $\delta E \gg \Gamma_{\rm el}$. 
In this regime
we can solve \Eq{eq:integraleq}
by iteration,  $\ff=\ff^{(0)}+\ff^{(1) } + \ldots$,  with $\ff^{(0)}(\h) = -2i \h/\delta E(h)$. 
Physically this expansion corresponds to the number of collisions, with  $\ff^{(1)}$ determining
the emission rate from one collision and so on. After straightforward algebra one 
finds\footnote{ Note that we have somewhat cavalierly shifted
   the integration variable 
   $\p_\perp \rightarrow \p_\perp + \q_\perp$ to re-write $\frac{\p_\perp^2}{\delta E^{2}(\p_\perp)}  \rightarrow \frac{1}{2}\left( \frac{\p_\perp^2}{\delta E^{2}(\p_\perp)} + \frac{(\p_\perp+\q_\perp)^2}{\delta E^{2}(\p_\perp+\q_\perp)} \right)$
   in order to write the integrand as a perfect square, which naturally appears in diagrammatic calculations of the single scattering rates~\cite{Ghiglieri:2015ala}.}
{
\st
\label{eq:BetheHeitlerTotal}
  \frac{\dd \Gamma_{\rm inel}^{\BH} (\p_0)}{\dd z} = 
2 \alpha_s  \,  P_{g\rightarrow g}^{\rm soft} (z) \, \int \frac{d^2p_\perp}{(2\pi)^2 }  \int d^2q_\perp \frac{d\Gamma_{\rm el}}{d^2q_\perp }
 \left( \frac{\p_\perp }{p_\perp^2 + m^2 } -  \frac{\p_\perp + \q_\perp }{(\p_\perp + \q_\perp)^2 + m^2  } \right)^2 \, . 
\stp
}
The large $\p_\perp$ limit  of this rate is known as the Gunion-Bertsch formula~\cite{Gunion:1981qs}
\st
  (2\pi)^2 \frac{\dd \Gamma_{\rm inel}^{\BH} (\Ps)}{\dd z \, \dd^2p_\perp} = 
2 \alpha_s  \,  P_{g\rightarrow g}^{\rm soft} (z) \, \frac{\hat q}{p_\perp^4 }  \, .
\stp
To estimate the total rate one can integrate this expression over $p_\perp$ down
to a scale  $p_\perp  \sim m$  yielding the Bethe-Heitler estimate given earlier in \Eq{bhestimate} .


In the opposite limit $\omega \gg \omega_{LPM}$  we can also find an approximate solution to \Eq{eq:integraleq}  known as the harmonic oscillator approximation. Since for $\omega \gg \omega_{LPM}$ the transverse momentum $h$ acquired over the formation time is large compared to the typical momentum transfer $q_{\perp}$ aquired in a single scattering $q_\perp \sim m$, one can expand the differences $\ff(\h) - \ff(\h + \q_\perp)$ for small $\q_\perp$, which transforms \eq{eq:integraleq} into a partial differential equation
\st
2 \h = i \delta E(h) \ff(\h)  -\frac{1+z^2+(1-z)^2}{8}  \, \hat q \, \delta^{ij}_\perp \frac{\partial^2}{\partial h^i \, \partial h^j }  \ff(\h)
\stp
By approximating $\delta E(h) \simeq \frac{h^2}{2P z(1-z)}$ and Fourier
transforming with respect to $\h$ (with ${\bm b}$ conjugate to $\h$), one
obtains a Schr\"odinger-like equation for a particle with an effective mass $M =
P z (1 -z)$ in an imaginary harmonic potential $V({\bm b}) = \tfrac{-i}{2} M
\omega_0^2  {b}^2$ with oscillation frequency $\omega_0^2=\hat{q}
\frac{1+z^2+(1-z)^2}{4 z (1-z) P}$. Solving this equation, one finds
that the final emission rate is proportional  to $\omega_0 \sim \Tform^{-1}$,
which in the soft limit $(z\ll1)$  yields
\begin{align}
   \label{eq:LPMemission}
 \frac{d\Gamma_{\rm inel}^{\LPM} (\Ps)}{dz} =&
 \frac{\alpha_s }{2\pi } \;  P_{g\rightarrow g}^{\rm soft}(z) \;  \sqrt{ \frac{\hat q}{ P z(1-z)}  } \, .
\end{align}
General expressions for the emission rates involving multiple species 
are given in \cite{Arnold:2008zu,Arnold:2008vd} in the same notation used here.
Comparing \Eq{eq:LPMemission} with the Bethe-Heitler limit of \Eq{bhestimate},
shows that the emission rate is controlled by the inverse of the formation time
$1/\Tform$ rather than the elastic scattering rate $\sim \hat q/m^2$ in
\Eq{bhestimate}, suppressing the emission of radiation at high energies.



%
%

\section{Basics of weak coupling thermalization}
\label{basics}

Now that we have outlined the basic physics of QCD kinetics, 
we will illustrate key features of the equilibration process 
in homogenous isotropic systems where a detailed understanding of the dynamics has been gained in a series of studies \cite{Kurkela:2011ti,Blaizot:2011xf,Berges:2013fga,Kurkela:2014tea}. 
Since the equilibration dynamics crucially depends on the properties of the initial state, it useful to
distinguish between systems which are initially far from equilibrium, and
systems which are initially close equilibrium. While in the latter case,
one expects a direct relaxation of the system to equilibrium
governed by an equilibrium rate, the situation is more complicated for systems
which are initially far from equilibrium, and various kinds of phenomena can
occur en-route towards thermal equilibrium. Nevertheless 
a general
characterization of the equilibration process can be achieved for broad classes
of far from equilibrium initial conditions. Specifically, for homogenous and isotropic systems one needs to distinguish between \emph{overoccupied} systems, i.e.
systems in which the energy is initially carried by a large number of low
energy degrees of freedom,  and \emph{underoccupied} systems, i.e. systems in which
the energy is carried by a small number of very high energy degrees of freedom. 
As we have emphasized in the previous section the first stage of
the bottom-up scenario corresponds to the ``over-occupied'' case, while the second and third stages
 correspond to the under-occuppied
case.

\subsection{Overoccupied systems}
\label{basics_overoccupied}

We first consider a system where the initial energy density 
is carried by a large number of low energy  degrees of freedom, i.e. if the
quasi-particle energy is $E_\p \sim Q$, then the energy density is $e  \sim
f_{0} Q^4$ where $f_0\gg1$ denotes the initial phase-space density.
Clearly, this initial state is very far from an equilibrium state, where
the energy density $e_{\rm eq} \sim T^4$ is carried by a smaller
number  of modes with $f \sim 1$ and $E_\p \sim T$.  Since energy is conserved during the evolution, 
the final temperatue $T \sim Q \, f_{0}^{1/4}$ at the end  of the equilibration process
is much larger than $Q$.
Because of  this large scale
separation between $Q$ and $T$,
the redistribution of energy from low energy modes
to high energy modes  is then
a classic problem of turbulence known as a direct energy cascade discussed in the next section~\cite{nazarenko2011wave}. 

\subsubsection{Non-thermal fixed points and the energy cascade} 

The initial evolution of
overoccupied plasmas can be equivalently described in
terms of classical fields or weakly interacting quasi particles,
due to an overlap in their respective range of validity
\cite{Mueller:2002gd,Aarts:1997kp,Jeon:2004dh}. 
For this reason
the initial evolution 
can either be studied using
classical-statistical simulations of the non-linear gauge field dynamics (see e.g. \cite{Berges:2013fga}), or using the numerical simulations and analytic considerations of  kinetic theory~\cite{Kurkela:2011ti,Blaizot:2011xf}.

It was found that the initial evolution of overoccupied systems 
proceeds  
via a quasi-stationary state referred to as a \emph{non-thermal fixed point} (NTFP).
Here the dynamics becomes insensitive
to the details of the initial conditions after a short time, and the evolution follows a
self-similar scaling behavior
\cite{Kurkela:2012hp,Schlichting:2012es,Berges:2013fga}. 
Indeed, the  phase-space density $f(t,\ps)$ in this regime evolves 
with the scaling form
\st
f(t,\ps) = (Qt)^{\alpha} f_{S}\left((Qt)^{\beta} \frac{p}{Q} \right)\;,
\label{overocuppied-form}
\stp
which is characteristic for non-stationary turbulent
processes~\cite{nazarenko2011wave} and  the scaling form in
\Eq{overocuppied-form} describes a \emph{direct energy cascade},
i.e. the transport of energy from low momentum to high momentum excitations
necessary to achieve thermalization. 

\begin{figure}
\includegraphics[width=4in]{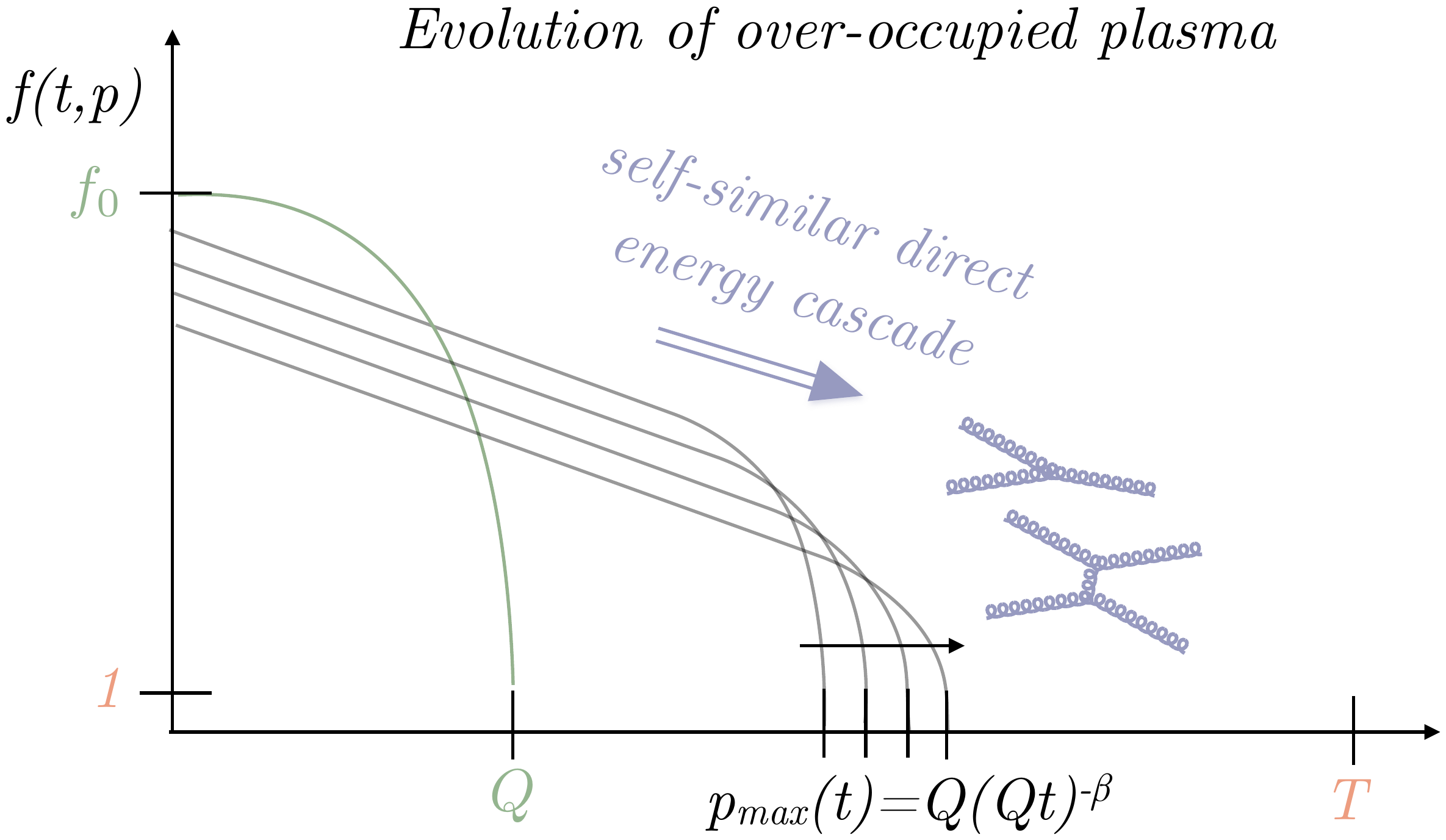}
\vskip 0.2in
\includegraphics[width=4in]{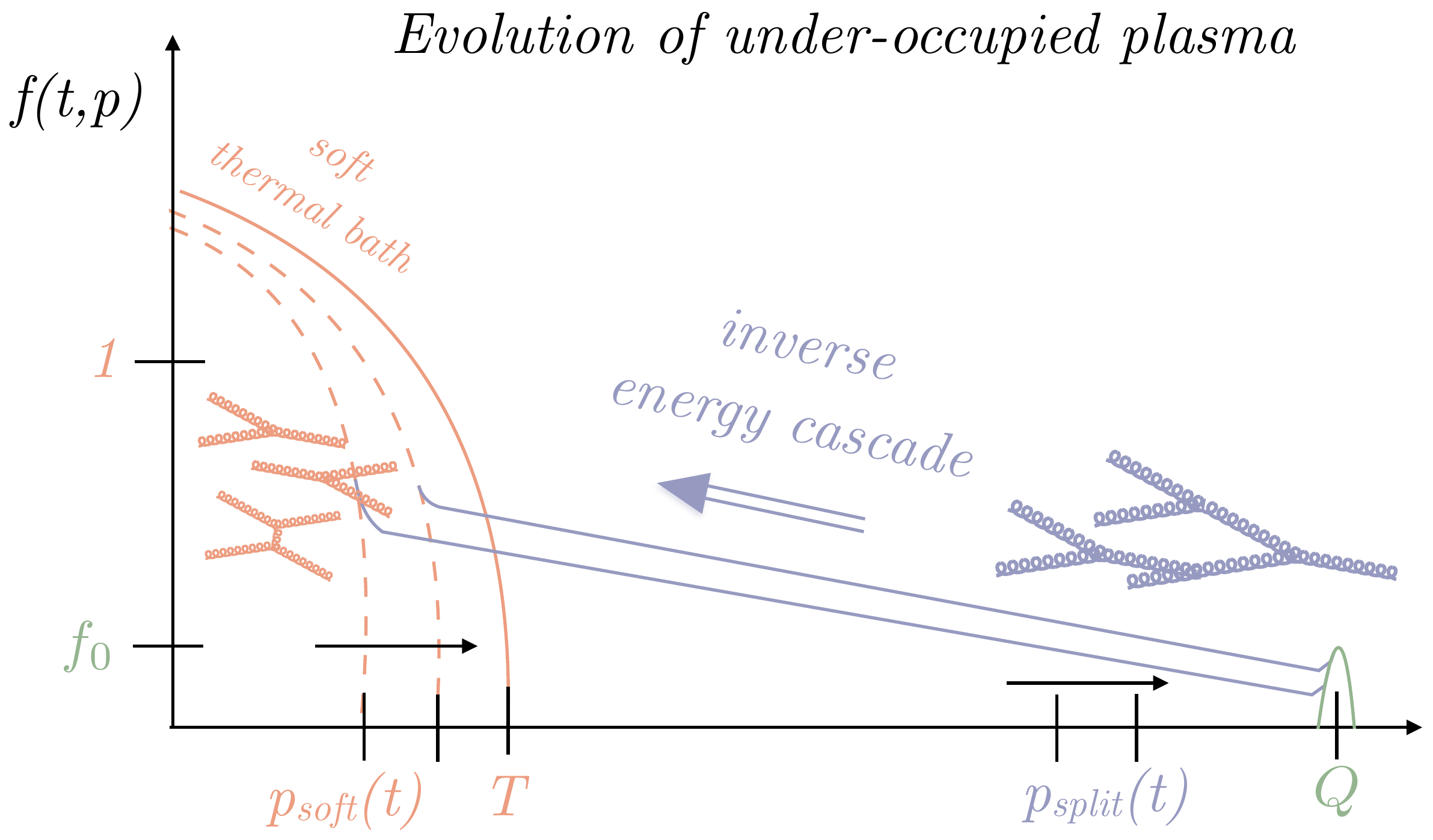}
\caption{\label{fig:OverUnderCartoon} Illustration of the thermalization process in over-occupied and under-occupied systems, summarizing the results of classical-statistical field simulations\cite{Schlichting:2012es,Kurkela:2012hp,Berges:2013fga} and kinetic theory simulations \cite{York:2014wja,Kurkela:2014tea}.}
\end{figure}


The scaling exponents $\alpha,\beta$ (which will be negative) determine the increase of
the characteristic momentum scale $p_{\rm max}(t) \sim Q (Qt)^{-\beta}$, and the
simultaneous decrease of the occupancy of hard excitations $f(t,p\sim p_{\rm
max}(t)) \sim (Qt)^{\alpha}$ (see \Fig{fig:OverUnderCartoon}). These scaling exponents can be determined from a straightforward scaling
analysis of the underlying kinetic equations \cite{Blaizot:2011xf,Kurkela:2011ti,Kurkela:2012hp,Berges:2013fga} following well established
techniques in the context of weak wave turbulence \cite{nazarenko2011wave}. One immediate
constraint on the scaling exponents $\alpha,\beta$ comes from the requirement
of energy conservation 
\st
e(t) = \int_{\ps} E_{\ps} \, f(t,\ps)=\textrm{const},
\stp
which for a self-similar evolution of the form in \Eq{overocuppied-form} gives rise to a
scaling relation
\begin{eqnarray}
\label{eq:EnergyConservationScaling}
\alpha-4\beta=0\;.
\end{eqnarray}
A second scaling relation can be inferred from a scaling analysis of the kinetic equation. Even though the full scaling analysis of all leading order kinetic processes is somewhat complicated, the essence can be understood by considering as an example small angle elastic processes, whose contribution to the collision integral in \Eq{eq:FPequation} is of the form of a Fokker-Planck equation,
where the drag coefficient  $\eta(t)$ and momentum diffusion coefficient $\hat{q}(t)$ are of order (see \Eq{FPcoefficients} and \Eq{eq:scalesmstar})
\begin{subequations}
   \label{eq:qhatestimate}
\begin{align}
   \eta(t) \sim&  \alpha^2_s \int_{\ps} \frac{f(t,\ps)}{p} \, ,  \\
   \hat{q}(t) \sim& \alpha^2_s \int_{\ps}f(t,\ps)\, (1+f(t,\p))  \, .
\end{align}
\end{subequations}
With the scaling ansatz of \Eq{overocuppied-form} in the 
high-occupancy regime $f(t,\p) \gg 1$,  these quantities scale (up to logarithmic corrections) as
\begin{align}
   \eta(t) \sim& (Qt)^{\alpha-2\beta}\; \alpha^2_s Q^2\int_{\qs} \frac{f_{S}(q)}{q}\;,  \\
\hat{q}(t) \sim& (Qt)^{2\alpha-3\beta}\;\alpha^2_s Q^3\int_{\qs} f_{S}^2(q)\;,
\end{align}
under the self-similar evolution of the system.  Based on this analysis one can establish a scaling behavior of the collision integral
\begin{eqnarray}
\label{eq:CollisionScaling}
C_{\rm diff}[f(t,\ps)] = (Qt)^{3\alpha-\beta} \, C_{\rm diff}[f_S(Q)]\;, 
\end{eqnarray}
which also extends to large angle elastic and inelastic processes \cite{Blaizot:2011xf,Kurkela:2011ti,Kurkela:2012hp,Berges:2013fga}. By matching the time dependence on the r.h.s of Eq.~(\ref{eq:CollisionScaling}) with that of the l.h.s. of the Boltzmann equation, one infers the dynamical scaling relation
\begin{eqnarray}
\alpha-1=3\alpha-\beta\;,
\end{eqnarray}
which along with Eq.~(\ref{eq:EnergyConservationScaling}) uniquely determines
the exponents. Strikingly, the scaling analysis of the kinetic equations also
reveals the universal nature of the dynamical scaling exponents, which are
insensitive to microscopic details of the underlying theory and take the values
$\alpha=-4/7$ and $\beta=-1/7$ for $SU(N_c)$ gauge theories in $d=3$ dimensions
\cite{Blaizot:2011xf,Kurkela:2011ti}. These are in line with classical-statistical field simulations of $SU(2)$ and
$SU(3)$ Yang-Mills plasmas \cite{Kurkela:2012hp,Berges:2013fga,Berges:2017igc}. 

Beyond the dynamics of energy transport, various perturbative and
non-perturbative properties of the NTFPs of $SU(N_c)$ Yang-Mills theory have
been investigated based on classical-statistical lattice simulations
\cite{Mace:2016svc,Berges:2017igc,Boguslavski:2018beu}  and show 
how the electric and magnetic sectors of kinetic theory emerge at late time.


\subsubsection{Equilibration} 
Eventually, the self-similar evolution breaks down 
when the energy has been transferred from the initial momentum scale $p_{\rm
max}(t=0)\sim Q$ all the way  
to the equilibrium
temperature $p_{\rm max}(t_{\rm eq})\sim T$~\cite{Blaizot:2011xf,Kurkela:2011ti}. 
Using the scaling exponent  $\beta$ and the initial occupancy $f_0 \sim
1/\alphas$, the self-similar cascade ends when
\begin{eqnarray}
   \label{turbulent_estimate}
t\sim t_{\rm eq} \sim  \alphas^{-2} f_{0}^{-1/4} Q^{-1} \sim \alphas^{-2} T^{-1} \, .
\end{eqnarray}
At the end of the cascade, the
phase-space occupancies of hard modes $f(t,p_{\rm max}(t))$ also becomes of
order unity, and the system is no longer parametrically far from equilibrium. 
The relevant scattering rates  
decrease over the course of the
cascade,
$\Gamma(t) \sim \hat{q}(t)\, p_{\rm
max}^{-2}(t) \sim \alphas^2 Q~(Qt)^{-1}$, and the final approach to equilibrium is ultimately controlled by
an equilibrium transport time scale, $\sim  \alphas^{-2} T^{-1}$. This time scale is
parametrically of the same order as the time scale for the turbulent transport
of energy given in \Eq{turbulent_estimate}. While the final approach to equilibrium is outside the range of
validity of classical-statistical simulations, it can be investigated further
based on numerical simulations in kinetic theory \cite{York:2014wja,Kurkela:2014tea}, which provide concrete,
rather than parametric, estimates of the thermalization time, $t_{\rm eq}\approx
0.46~\alphas^{-2} N_{c}^{-2} T^{-1}$~\cite{Kurkela:2014tea}.



\subsection{Underoccupied systems}
\label{basics_underoccupied}
We now consider the opposite case where  the initial energy density $e \sim Q^4
$ is carried by a small number $f_{0} \ll 1$ of high energy
degrees of freedom with $E_\p \sim Q$, and note that  this setup is 
reminiscent of a high-energy jet carrying a significant fraction of the energy
of the system. While the final equilibrium temperature can again be
inferred using energy conservation as $T \sim f_{0}^{1/4} Q$, the hierarchy of scales is now inverted with $T
\ll Q$. Since the equilibrium temperature $T$ is
much smaller than the characteristic momentum scale $Q$, the thermalization
process now requires a re-distribution of energy from high energy to low energy   degrees of freedom. 

Eventually the re-distribution of energy is achieved by an \emph{inverse energy
cascade} through multiple radiative branchings of the high energy particles~\cite{Baier:2000sb,Kurkela:2011ti,Blaizot:2013hx}.
However, 
before the inverse cascade 
can be established, a small fraction of the energy must be
transferred to low energy modes by direct emission of soft radiation. As
discussed in \Sect{sect_bath}, these
low energy modes thermalize quickly, creating of a soft
thermal bath and setting the 
stage for the inverse energy cascade described in \Sect{sect_cascade}.

\subsubsection{Direct radiation and creation of soft thermal bath}
\label{sect_bath}

Let us analyze how the soft bath is created. Following \cite{Kurkela:2011ti} there is
a competition between direct radiation from the hard modes, which tends to populate the
soft bath, and momentum diffusion which tends to push the typical
momentum scale of the bath to higher momentum.
As we will show below,  direct radiation initially dominates and over populates the bath. 
Then,  as the LPM effect sets in and suppresses additional
radiation, the soft bath reaches an occupancy of order unity with an 
equilibrium temperature $T_{\rm soft}(t)$. 

Initially, elastic scattering processes amongst the hard modes
occur relatively frequently,  with a rate of  order 
\[
   \Gamma_{\rm el} \sim  \frac{\hat{q}}{m^2} \sim \alphas Q \, ,
\]
where we have estimated $\hat q$ and $m$ from
the distribution of hard particles 
\begin{subequations}
   \label{eq:hardqhat}
\begin{align}
   \hat{q}_{\rm hard} \sim& \alphas^{2} \int_\ps f_\ps(1 + f_\ps) \sim  \alphas^2 f_{0} Q^3 \,,  \\
   m_{\rm hard}^2 \sim&  \alphas \int_\ps \frac{f_\ps}{p} \sim \alphas f_{0} Q^2 \, .
\end{align}
\end{subequations}
These elastic scatterings  induce soft and collinear radiation,  and it is these processes which are responsible  for creating the soft bath.
%
%
From the first line of \Eq{C12}, the rate at which soft particles with momentum $\p$ are produced 
by the  hard particles with momentum $\Ps \sim Q$ is initially 
\begin{eqnarray}
   \frac{\partial f(t,\ps)}{\partial t}\simeq \nu_g \int_{\Ps} \int_{0}^{1} dz~ \frac{\dd \Gamma_{\rm inel}(\Ps)}{\dd z} ~\frac{(2\pi)^3}{\nu_g} \delta^{(3)}(\ps -z\Ps)~ f_0(\Ps)~\Big(1+ f_0(\Ps)\Big)\;. 
\end{eqnarray}
Note that this rate is independent of the soft phase space density $f(t,\p)$ 
due to a cancellation between the gain and loss terms \cite{Kurkela:2011ti}.

The radiated soft fragments are of course more susceptible to elastic
scattering processes, and have the chance to  equilibrate via both elastic
scatterings and inelastic processes, giving rise to a
dynamical scale 
\begin{eqnarray}
\psoft(t) \sim \sqrt{\hat{q}(t) t} \sim \alphas f_{0}^{1/2} Q (Qt)^{1/2} \, .
\end{eqnarray}
Soft fragments 
below $\psoft(t)$
have an effective temperature $T^{*}_{\rm soft}(t)$ (defined precisely below) characterizing the occupancy of these modes.


As we will now estimate, the phase space densities become initially overoccuppied as the soft bath 
is built up.  
This happens because at early times the particles are copiously produced via 
Bethe-Heitler radiation, and do not have time to increase $\psoft$ through diffusion.
The Bethe-Heitler approximation is appropriate here because $\psoft \ll \omega_{\LPM}$ as discussed in \Sect{sec:collinear_rad}.  
The occupancy of the soft sector 
can be estimated from the amount of energy $e_{\rm soft}$ radiated
into this sector and $\psoft(t)$. The radiated energy is of order
\begin{eqnarray}
   \label{eq:radiatedenergy}
   e_{\rm soft}(t) \sim \int_0^{t} dt~\int_{\ps}^{p_{\rm max}(t)}
E_{\ps} \frac{\partial f(t,|\ps|)}{\partial t} \sim e_{\rm hard}
 \int_{0}^{\psoft(t)/Q}dz~z
\frac{\dd \Gamma_{\rm inel}(Q)}{\dd z}  t \, , 
\end{eqnarray}
which, with the Bethe-Heitler estimate for $\dd\Gamma_{\rm inel}^{\BH}/\dd z$ from \Eq{bhestimate}, yields
\st
e_{\rm soft}(t) \sim \alpha_s \, e_{\rm hard} \,  \frac{(\hat q(t) t)}{m^2} \frac{\psoft(t)}{Q} \, .
\stp
Using the estimates for $\hat q$ and $m$ in
\Eq{eq:hardqhat}, 
one finds that the  effective temperature of the soft sector is
given by 
\st
T^{*}_{\rm soft}(t) \equiv  \frac{e_{\rm soft}(t)}{ (\psoft(t))^3  }  \sim Q \, ,   
\stp
and thus, since $T^{*}_{\rm soft}(t)$ is much larger than the 
characteristic momentum scale $\psoft(t)$,
the system is initially  over occupied for a short period of time
\st
   (Qt)^{1/2} \lesssim \alphas^{-1} f_{0}^{-1/2} \, .  
\stp


The radiated soft excitations
will ultimately contribute to screening and scattering processes. 
While at early times these contributions are negligible, their contributions increase as a function of time according to
\begin{subequations}
   \label{eq:qhatsoft}
\begin{eqnarray}
m_{\rm soft}^2(t)\sim \alphas \, T^{*}_{\rm soft}(t)\, \psoft(t)
\sim m_{\rm hard}^2 \, \frac{(Qt)^{1/2}}{f_{0}^{-1/2} \alphas^{-1}} \;,  \\
\hat{q}_{\rm soft}(t) \sim \alphas^2 \, (T^{*}_{\rm soft}(t))^2 \, \psoft(t) \sim \hat{q}_{\rm hard} \, \frac{(Qt)^{1/2}}{f_{0}^{-1/2} \alphas^{-1}}\;,
\end{eqnarray}
\end{subequations}
and thus for $(Qt)^{1/2} \gtrsim \alphas^{-1} f_{0}^{-1/2}$ they become of the same order as the contributions from the hard sector, and the systems enters the second stage of the thermalization process.

For $(Qt)^{1/2} \gtrsim \alphas^{-1} f_{0}^{-1/2}$, the radiative dynamics continues in a similar fashion, but now the soft and
hard sectors now give comparable contributions to elastic scattering, while the  screening is
 dominated by the soft sector.
The emission of soft radiation at the
characteristic scale $\psoft(t)$ now suffers from LPM
suppression as now $\psoft(t)$ has become of order  of
$\omega_{\LPM}$. Substituting \Eq{eq:LPMemission} in \Eq{eq:radiatedenergy},
the amount of energy radiated directly into
soft modes $p\sim \psoft(t)$ is now given by
\begin{eqnarray}
   \label{eq:epsilonsoftstage2}
e_{\rm soft}(t) \sim \alphas e_{\rm hard} \sqrt{\frac{\hat{q}(t) t^2}{Q}} \sqrt{\frac{\psoft(t)}{Q}}\;,
\end{eqnarray}
which along with the consistency relations
\[
e_{\rm soft} \sim  T^{*}_{\rm soft}(t)  \psoft^3(t)\,, \qquad  
\psoft(t) \sim \sqrt{\hat{q}(t) t} \,,  \qquad \hat{q}(t) \sim \hat{q}_{\rm
soft}(t) \sim \alphas^2 T^{*}_{\rm soft}(t)^2 \psoft(t) \, ,
\]
determines the dynamical evolution of the soft sector. One finds that the
characteristic momentum scale $\psoft(t)$ continues to
increase, while the effective temperature $T^{*}_{\rm soft}(t)$ of the soft
sector drops
\begin{align}
   \psoft(t) \sim & \alphas f_0^{1/2}~Q~(Qt)^{1/2}\;,   \\
   T^{*}_{\rm soft}(t)\sim& \alphas^{-1/2} f_{0}^{1/4}~Q~(Qt)^{-1/4}\;.
\end{align}

Eventually, at a time $Qt \sim f_{0}^{-1/3} \alphas^{-2}$ the characteristic
momentum scale $\psoft(t)$ becomes comparable to the effective
temperature $T^{*}_{\rm soft}(t)$, indicating that the phase-space
densities of soft particles $f(\psoft(t)) \sim 1$ are now
of order unity, and the soft sector can be considered thermalized from now on.
At this time only a small fraction $e_{\rm soft} \sim f_{0}^{1/3}
e_{\rm hard}$ of the energy $e_{\rm hard}$ of the hard particles has been
transferred to the soft thermal bath via direct radiation. 

\subsubsection{Inverse energy cascade}
\label{sect_cascade}
In addition to directly radiating  soft gluons with $p\lesssim \psoft(t)$, 
the hard modes can transfer energy to soft sector via multiple
successive branchings. Although soft branchings with ${\rm min}(z,1-z) \ll 1$
occur most frequently, 
quasi-democratic branchings with
$z\sim 1/2$ are more efficient in transferring energy, 
and this will give the dominant contribution to energy transport at late times. Because of the
characteristic energy dependence of the LPM splitting rates in \Eq{eq:LPMemission}, there is a  momentum   scale 
\begin{eqnarray}
   \label{eq:psplit}
p_{\rm split}(t) \sim \alphas^2 \, \hat{q}(t) \,  t^2\;, 
\end{eqnarray}
where the probability $t \, \dd \Gamma(p_{\rm split}(t))/\dd z$ to undergo a
quasi-democratic splitting with $z\sim 1/2$ is of order unity. 
$p_{\rm split}(t)$ is the momentum of the most energetic particles that can be stopped over  
the total lifetime $t$ of the system~\cite{Arnold:2009ik}.
Clearly at early
times $p_{\rm split}(t) \ll \psoft(t)$, and quasi-democractic branchings
contribute very little to the overall energy transfer.
However,  at a time of order $Qt
\sim f_{0}^{-1/3} \alphas^{-2}$, 
$p_{\rm split}(t)$ becomes of order of $\psoft(t)$, 
and multiple successive branchings begin to dominate the energy transfer 
to the soft thermal medium.  Hence the last stage of the thermalization process
is analogous to  a highly energetic jet loosing
energy to the QGP, highlighting an important connection between jet
quenching and thermalization. 

In the final stages the soft bath is equilibrated, and $\hat q(t)$ and $e_{\rm soft}(t)$ are determined by their equilibrium values at temperature $T_{\rm soft}(t)$, wich depends on time
\begin{eqnarray}
   \label{equilibriumqhat}
   \hat{q}(t) \sim \hat{q}_{\rm soft}(t) \sim \alphas^2 (T_{\rm soft}(t))^3 \;, \qquad e_{\rm soft}(t) \sim (T_{\rm soft}(t))^4 \, . 
\end{eqnarray}
To determine the rate of energy transfer,  we need to compute the energy radiated up to the  momentum scale $p_{\rm split}(t)$, 
which  will then have time enough to undergo successive branchings in the bath.
Using  \eqref{eq:radiatedenergy} with the LPM estimate for $\dd\Gamma/\dd z$  from \eqref{eq:LPMemission},
the transfer of energy from hard to soft modes is of order
\begin{eqnarray}
e_{\rm soft}(t)  \sim \int_0^t dt~\int_{\ps}^{p_{\rm split}(t)} E_{\ps} \frac{\partial f(t,|\ps|)}{\partial t} \sim \alphas e_{\rm hard} \sqrt{\frac{\hat{q}(t) t^2}{Q}} \sqrt{\frac{p_{\rm split}(t)}{Q}} \, ,
\end{eqnarray}
yielding with \eqref{eq:psplit} the estimate
\begin{eqnarray}
   \label{eq:esoftlast}
e_{\rm soft}(t) \sim e_{\rm hard}~\frac{p_{\rm split}(t)}{Q} \, .
\end{eqnarray}

The transfer of energy ends when the thermal medium has
entirely absorbed the energy of the hard partons $e_{\rm soft}(t) \sim e_{\rm hard}$
which occurs when  $p_{\rm split}(t)\sim Q$. 
Self-consistently 
determining the time evolution of the scales according to \Eq{eq:esoftlast}
and \eqref{equilibriumqhat}, we find 
 $p_{\rm split}(t)\sim \alphas^{16} f_{0}^{3} Q (Qt)^{8}$ and $T_{\rm soft}(t)\sim \alphas^{4} f_{0} Q (Qt)^{2}$, thus, at a time of order
\begin{eqnarray}
t_{\rm eq} \sim \alphas^{-2} f_{0}^{-3/8} Q^{-1} \sim  \alphas^{-2} T^{-1} \sqrt{\frac{Q}{T}} \, ,
\end{eqnarray}
the temperature of the soft thermal bath $T_{\rm soft}(t)$ becomes of the order
of the final equilibrium temperature $T \sim f_{0}^{1/4} Q$.
In
contrast to the overoccupied case,
the equilibration time of an underoccupied system $t_{\rm eq} \sim  \alphas^{-2}
T^{-1} \sqrt{Q/T}$ is parameterically larger than the near-equilibrium
relaxation rate $\sim \alphas^{-2} T$. Notably, the additional dependence on the ratio of momentum
scales $\sqrt{Q/T}$ implies that excitations with different energies $Q$
equilibrate on different time scales. 

Beyond the level of parametric estimates  \cite{Kurkela:2011ti} a more quantitative description of the inverse energy cascade has been put forward already in the original bottom-up paper \cite{Baier:2000sb}; the connections to wave turbulence were established in subsequent works \cite{Blaizot:2013hx,Blaizot:2015jea} in the context of jet quenching. Within an inertial range of momenta $T_{\rm soft}(t) \ll |\ps| \ll Q$ the dynamics is dominated by successive branchings, as described by an effective kinetic equation of the form
\begin{eqnarray}
   \label{cascadeboltz}
   \frac{\partial}{\partial t} f(t,|\ps|) \simeq \int_{0}^{1}dz \left[z^{-3} \,  
   \frac{\dd \Gamma_{\rm inel}^{\LPM}(\p/z)}{\dd z} ~f\Big(t,\frac{\ps}{z}\Big)  - \frac{1}{2} \frac{\dd \Gamma_{\rm inel}^\LPM(\ps)}{\dd z} f(t,\ps)\right] \, ,
\end{eqnarray}
By exploiting the symmetry $\dd\Gamma_{\rm inel}^{\LPM}(\ps,z) =
\dd\Gamma_{\rm inel}^{\LPM}\Big(\ps,1-z\Big)$, and the approximate scale invariance
of the splitting rates $\dd\Gamma_{\rm inel}^{\LPM}(\ps/z,z)
\simeq \sqrt{z}~\dd\Gamma_{\rm inel}^{\LPM}(\ps,z)$, the collision integral
in \eqref{cascadeboltz} an be transformed into 
\begin{eqnarray}
\label{eq:ZhakTrafoCascade}
\frac{\partial}{\partial t} f(t,\ps) \simeq \int_{0}^{1}dz~ \frac{\dd \Gamma_{\rm inel}^{\LPM}(\ps) }{\dd z}  \left[ z^{-5/2} f\Big(t,\frac{\ps}{z}\Big)  - z f(t,\ps)\right]\;.
\end{eqnarray}
Eq.~\eqref{eq:ZhakTrafoCascade} admits stationary solutions of the Kolmogorov-Zakharov form 
\begin{eqnarray}
\label{eq:KZSpec}
f_{KZ}\Big(t,\ps\Big) = f^{*}~\left(\frac{Q}{|\ps|}\right)^{\kappa} \,, 
\end{eqnarray}
with a universal spectral index $\kappa=7/2$ and a non-universal amplitude $f^{*}$. One finds that -- in analogy to the Kolmogorov-Zakharov spectra of weak wave turbulence -- the solution is associated with scale independent energy flux, meaning that the energy
lost by modes above a scale $\Lambda$
\begin{eqnarray}
\label{eq:EnergyFluxDef}
\frac{d}{dt} e_{\rm hard}(t) \simeq \int_{\Lambda}^{\infty} 4\pi p^2 dp \,  E_\ps~\frac{\partial}{\partial t} f(t,\ps)\;,
\end{eqnarray}
is independent of $\Lambda$. This property reflects the transport of  energy from hard
modes $(\Lambda \sim Q)$ all the way to the soft thermal bath $(\Lambda \sim
T_{\rm soft}(t))$ via successive branchings. By exploiting the scale invariance
of the collision integral $\dd\Gamma_{\rm inel}^{\LPM}(\ps,z) \simeq
\sqrt{Q/|\ps|} \,  \dd\Gamma_{\rm inel}^{\LPM}(Q,z)$, the energy flux in
\Eq{eq:EnergyFluxDef} can be evaluated  
by  using
\Eq{eq:ZhakTrafoCascade} and \Eq{eq:KZSpec} to evaluate $\partial_t f$, 
and by taking the limit where the
spectral exponent  approaches the Kolmogorov-Zakharov
solution from above~\cite{nazarenko2011wave,zakharov2012kolmogorov}, $\kappa \searrow 7/2$. This  yields
\begin{eqnarray}
   \frac{d}{dt} e_{\rm hard}(t) \simeq -(4\pi)~Q^5~f^{*}~\gamma_{\rm g}\;, \qquad \gamma_{g}=Q^{-1} \int_{0}^{1} dz~ \frac{\dd\Gamma_{\rm inel}^{\LPM}(Q,z)}{\dd z} ~z\log(1/z)\;.
\end{eqnarray}
While the inverse energy cascade is ultimately responsible for transferring the
energy of hard particles to the soft bath, coincidentally the properties of the
QCD splitting functions are such that a single emission is sufficient to create
the turbulent spectrum \Eq{eq:KZSpec} within the inertial range of momenta
$T_{\rm soft}(t) \ll |\ps| \ll Q$
\cite{Baier:2000sb,Blaizot:2013hx,Blaizot:2015jea,Mehtar-Tani:2018zba}. Based
on this peculiar property, it is then also possible to estimate the amount of
energy injected into the cascade (corresponding to the non-universal amplitude
$f^{*}$) and calculate the energy transfer to the thermal bath as discussed in
detail in \cite{Baier:2000sb,Mehtar-Tani:2018zba}. We also note that numerical
studies of the thermalization of underoccupied systems performed in
\cite{Kurkela:2014tea} confirm the basic picture of the thermalization
mechanism illustrated in Fig.~\ref{fig:OverUnderCartoon} and provide additional
information on the thermalization time.

\subsection{Generalization to anisotropic systems}
\label{bupwnumbers}

So far we have discussed the thermalization process for statistically isotropic
plasmas. When the distribution is anisotropic, a quantitative analysis of the
evolution becomes significantly more complicated due to the presence of plasma
instabilities~\cite{Mrowczynski:1993qm,Romatschke:2003ms}. 
Once the phase space distribution has an order one
anisotropy, instabilities
qualitatively change the screening mechanisms in
the plasma, and significantly complicate the calculation of radiation rates and
the relaxation to equilibrium~\cite{Arnold:2003rq}. 
How precisely plasma instabilities modify the
thermalization process in over-occupied and under-occupied systems has not been
fully clarified, although a number of proposals exist~\cite{Kurkela:2011ub,Bodeker:2005nv}.  However, it is known that such instabilities are much
\emph{less} important than in QED plasmas since the non-linear non-abelian
character of the field equations ultimately limits the growth of the instability
~\cite{Rebhan:2004ur,Arnold:2005vb}. 
While for overoccupied systems first-principles studies  including the dynamics of instabilities could be performed with classical-statistical field simulations, these simulations are technically challenging, and most studies in this context have focused on the
growth of instabilities at very early times. Since the situation remains
somewhat inconclusive -- especially with regards to underoccupied systems where
classical-statistical simulations are inapplicable -- we will ignore the
effects of plasma instabilities throughout the remainder of this section, and
only comment on selected results in our outline of the original bottom-up picture.
Current implementations in kinetic theory also have ignored plasma
instabilities to date~\cite{Kurkela:2015qoa}.

\begin{figure}
\begin{center}
\begin{minipage}{3in}
\includegraphics[width=3in]{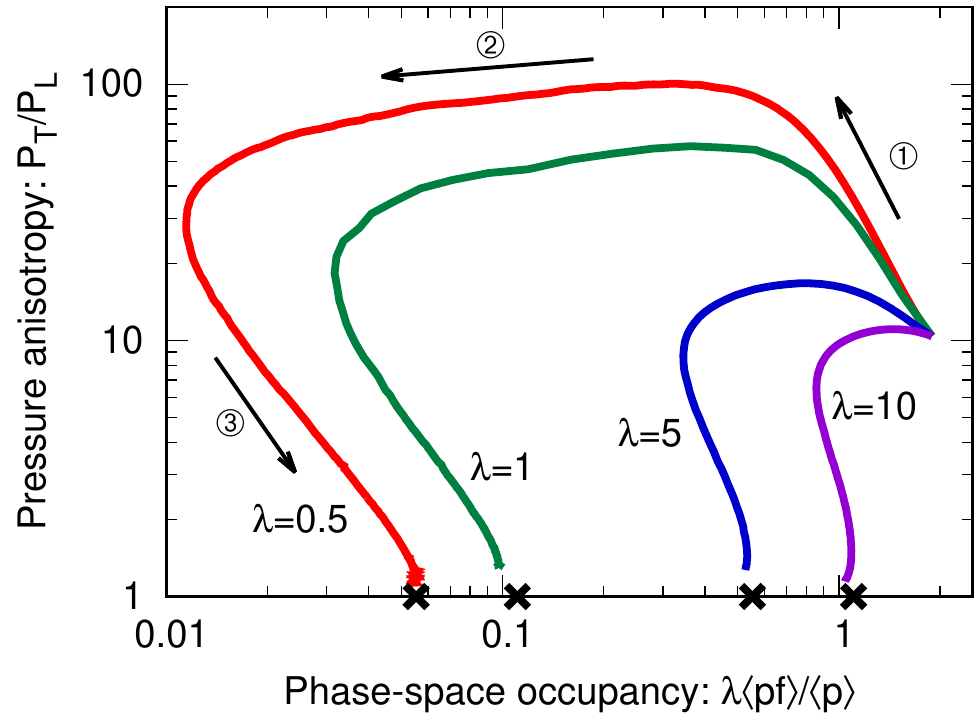}
\end{minipage}
\end{center}
\caption{\label{fig:BottomUpSimulations} Kinetic theory simulation of the non-equilibrium evolution of the pressure anisotropy and phase-space occupancy (see \Eq{occupancydefined}  and surrounding text) for a pure Yang-Mills
plasma in the original bottom-up scenario~\cite{Kurkela:2015qoa}. Here
$\lambda=4\pi\alphas N_c$ is the coupling,  and the black crosses indicate equilibrium value. The three arrows and associated circled numbers indicate the three stages of bottom-up.}
\end{figure}

As discussed qualitatively in \Sect{Bottomup}, 
the first and second/third stages of the bottom-up  scenario
are characteristic of overoccuppied and underoccupied systems respectively.
In all stages the presence of the longitudinal expansion  modifies the rates
discussed in \Sect{basics_overoccupied} and \Sect{basics_underoccupied} for static systems, without changing the
overall picture. 
In \Fig{fig:BottomUpSimulations}  we show a simulation result of Kurkela and Zhu of the original
bottom-up scenario~\cite{Kurkela:2015qoa}.
The simulation uses the 't Hooft coupling  $\lambda = 4 \pi \alphas N_c$ 
(and thus a ``realistic'' coupling
is $\lambda\simeq 10$ or more\footnote{In terms of macroscopic properties, the shear viscosity of the simulation is $\eta/s\simeq 0.62$ for $\lambda=10$.}),
and starts from a CGC motivated initial
condition characterized by 
\st
\label{KurkelaZhuIC} 
\frac{1}{\nu_g} \frac{dN}{d^2x_\perp dy}=0.23 \,\frac{Q_s^2}{\lambda}\,, \qquad   \sqrt{\llangle p_T^2 \rrangle  } = 1.8 \, Q_s,
\stp
  treating screening with one overall mass $m^2$ given by \Eq{eq:scalesmstar}. 
%
The pressure anisotropy is defined from the stress tensor $P_T/P_L\equiv(T^{xx} + T^{yy})/(2 T^{zz})$,
while the occupancy in units of $\lambda^{-1}$ is
\st
\label{occupancydefined}
      \frac{\lambda \llangle p f_\p \rrangle}{\llangle p  \rrangle} = \frac{\lambda \int_\ps |\ps| \, f^2_\ps  }{\int_\p |\ps|  f_\ps } \, , 
\stp
which in equilibrium reaches $0.11\lambda$, indicated by the crosses in \Fig{fig:BottomUpSimulations}.

The numerical simulations confirm the three stage picture of bottom up
thermalization:  in stage one the the anisotropy grows and the occupancy decreases; in stage two the occupancy decreases and the anisotropy is stabilized; and finally in stage three the anisotropy 
approaches unity and  the energy of the system is thermalized.  In the weak
coupling limit ($\lambda\simeq 0.5$) the three
different stages are clearly visible, whereas for more realistic coupling
strength $(\lambda \simeq 10)$ the distinctions between the different stages
becomes increasingly washed out.
We will describe each stage more completely below
using the results of \Sect{basics_overoccupied} and \Sect{basics_underoccupied}.

To analyze the first over-occupied stage in the expanding case
we examine the Boltzmann equation with an elastic scattering
\st
\label{Expandingfp} 
\left( \frac{\partial}{\partial \tau} - \frac{p_z }{\tau} \frac{\partial }{\partial p_z  }\right) f(\tau, p_z,\p_\perp) = \frac{\hat q }{4}  \, \frac{\partial^2 f}{\partial p_z^2 } \, .
\stp
Here the free streaming term  on the l.h.s. stems from the expansion 
of the system, and makes the momentum distribution increasingly anisotropic~\cite{Baym:1984np}.
On the r.h.s. is the Fokker-Planck operator discussed in \Sect{basics_kinetics}, 
but here we have kept only the most relevant term which competes with the
expansion and broadens the momentum distribution.  Eq.~\eqref{Expandingfp} admits a scaling solution of the form
\st
\label{NTFPbup}
f(\tau, p_z, p_T) 
= \frac{1}{\alpha (Q_s \tau)^{2/3} } f_S \left(\frac{p_T } {Q_s}  ,  
\frac{p_z (Q_s\tau)^{1/3}  } {Q_s} \right) \, ,
\stp
provided one uses the by now familiar estimate for $\hat q$ dominated
by the hard modes, $\hat q \sim \alpha^2 \int_\p f_\p (1 + f_\p)$. 
This scaling solution,  which features a decreasing  occupancy and an increasing anisotropy, is clearly seen in the numerical simulations of \cite{Kurkela:2015qoa} at least for the  smallest couplings. 

The first over-occupied stage of the bottom up scenario has also been
addressed in detail within classical-statistical simulations~\cite{Berges:2013fga,Berges:2013eia,Berges:2017igc}.
It was found that the phase space distribution in the classical simulations reaches the universal scaling form of \Eq{NTFPbup}, reflecting the NTFP discussed in \Sect{basics_overoccupied}.
In these simulations the effects of plasma instabilities are clearly observed at early times
during the approach to the NTFP attractor, but do not appear to
significantly affect the longitudinal momentum broadening in the scaling
regime, such that $\langle p_{z}^{2} \rangle \sim Q_s^2 (Q_s \tau)^{-2/3}$
decreases at late times as in the original bottom scenario.  It remains an open question 
why plasma instabilities do not seem to play a more important role during 
the first phase of bottom up.

From the scaling solution in \Eq{NTFPbup}, we see that the first phase ends at
a time $Q_s\tau \sim \alpha_s^{-3/2}$, and after this point the system in is an under-occupied non-equilibrium state. The estimates  and physics
for the thermalization of such states described in \Sect{basics_underoccupied} 
can be adapted to the expanding case by  recognizing
that hard modes are essentially free streaming, and thus the energy and number 
densities of these modes are continually decreasing, 
so that  the energy and number per rapidity ($\tau e$ and $\tau n$ respectively)
remains fixed:
   \begin{align}
         \tau e_{\rm hard}(\tau) =&   \frac{Q_s^3 }{\alpha_s}\, , \\
          \tau n_{\rm hard}(\tau) =&   \frac{Q_s^2}{\alpha_s}\, .
\end{align}
Using the estimate
\st
\label{eq:qhathardstage2}
   \hat q(\tau) \sim \alpha_s^2 \int_{\ps} f_\ps (1 + f_{\ps}) \sim \alpha_s^2\,  n_{\hard}(\tau)  \, ,
\stp
one finds that  because of the expansion the soft scale $p_{\rm soft}(\tau)$ remains constant in time
\st
\psoft^2(\tau) \sim \hat q(\tau) \tau \sim \alpha_s Q_s^2 \,  ,
\stp
as opposed to increasing as it does in the non-expanding case.  Thus, the 
pressure anisotropy in the second phase is constant and large as seen
in \Fig{fig:BottomUpSimulations}.
Eq.~\eqref{eq:epsilonsoftstage2} for the energy density produced by 
direction radiation by the bath into the soft modes remains valid
\begin{equation*}
   e_{\rm soft}(\tau) \sim \alphas \tau e_{\rm hard}(\tau) \sqrt{\frac{\hat{q}(\tau) }{Q_s}} \sqrt{\frac{\psoft(\tau)}{Q_s}}\;,\tag{\ref{eq:epsilonsoftstage2}}
\end{equation*}
but now $e_{\rm hard}(\tau)$ and $\hat q(\tau)$ are functions of time.
Qualitatively, Eq.~\ref{eq:epsilonsoftstage2} will hold 
even if plasma instabilities are present, but $\hat q$ will  
deviate from the estimate in \Eq{eq:qhathardstage2}, which is based upon
elastic scattering by the hard modes. However, because the 
plasma instabilities are bounded they will not radically change the picture.
The second phase of bottom-up ends when $e_{\rm soft}(\tau) \sim \psoft^4(\tau)$ and the soft bath has thermalized.  
Equating these two expression one finds that the second phase ends at
a time of order $Q_s \tau \sim \alpha_s^{-5/2}$.

Finally,  we analyze the last phase of bottom-up. Here again the physics is identical to the inverse energy cascade discussed in detail in \Sect{sect_cascade} for the static system.
Eq.~\eqref{eq:psplit} for the splitting (or stopping) momentum $p_{\rm
split}(\tau)=\alphas^2 \hat q(\tau) \tau^2$, 
and \Eq{eq:esoftlast} for $e_{\rm soft}(\tau)$  
are unchanged 
\begin{equation*}
   e_{\rm soft}(\tau) \sim e_{\rm hard}(\tau)~\frac{p_{\rm split}(\tau)}{Q_s} \, ,
\tag{\ref{eq:esoftlast}}
\end{equation*}
provided the free streaming result for $e_{\rm hard}(\tau)$ is used. 
Again, plasma instabilities
may modify our estimate for $\hat q(\tau)$, but this 
will not change the overall picture.
The system is completely thermalized when $e_{\rm soft}(\tau)$ becomes
comparable to $e_{\rm hard}(\tau)$,
$\tau e_{\rm soft}(\tau) \sim \tau e_{\rm hard}(\tau)  \sim Q_s^3/\alpha$. 
Using the fact that  $\hat q$ is determined by $e_{\rm soft}$ in equilibrium,
$\hat  q(\tau) \sim \alphas^2 \, e^{3/4}_{\rm soft}(\tau)$,  
one readily establishes 
that the system thermalizes at 
\st
     Q_s \tau \sim   \alpha_s^{-13/5} \, .
\stp

We hope that it is evident that the overall picture of bottom-up 
is quite robust. Ultimately this picture follows from a hard scale $Q_s$, kinematics, 
and generic features of collinear radiation. These features tend
to fill up a soft sector first, which then causes a cascade of 
the energy of the system to the IR. Indeed, an extensive analysis
of thermalization when plasma instabilities are present finds
many of the same qualitative features of bottom-up with somewhat modified
exponents~\cite{Kurkela:2011ub}.



%


\section{Simulations of early time dynamics and heavy-ion phenomenology}
\label{sec:pheno}

\subsection{Approach to hydrodynamics}
\label{sec:hydrodynamization}
We now turn to simulations of the early time dynamics and the approach to equilibrium
in high-energy heavy ion collisions. 
Here we will focus on the eventual
approach towards local thermal equilibrium,  and
determine when the evolution
can be described  with
relativistic viscous fluid dynamics.  

Viscous fluid dynamics describes the macroscopic evolution of the energy-momentum tensor $T^{\mu\nu}$, based on an expansion around local thermal
equilibrium which is controlled by the Knudsen number\footnote{${\mathcal
   T}_{\rm macro}$ is a typical macro timescale, which can be estimated from the
inverse of expansion scalar $(\nabla \cdot u)\equiv \mathcal T_{\rm macro}^{-1}$ of the fluid. For a Bjorken expansion $\mathcal T_{\rm macro} = \tau$.}  ${\rm Kn}_{\theta} \sim
\tau_{\rm micro}/{\mathcal T}_{\rm macro}$, 
   and the proximity to the equilibrium  state, which 
can be quantified
 by the 
non-equilibrium corrections to the stress tensor, $T^{\mu\nu}_{\rm non-eq}/T^{\mu\nu}_{\rm eq}$. 
At early times $\tau \sim 1/Q_s$ the longitudinal pressure
is  much smaller than the transverse pressure $P_L\ll P_T$
and hydrodynamics does not apply.
Consequently, the key question is to understand how
$T^{\mu\nu}$ then evolves towards  local thermal equilibrium where the longitudinal and transverse pressures are equal $P_L=P_T$.

\begin{figure}
\begin{minipage}{2.5in}
\includegraphics[width=2.5in]{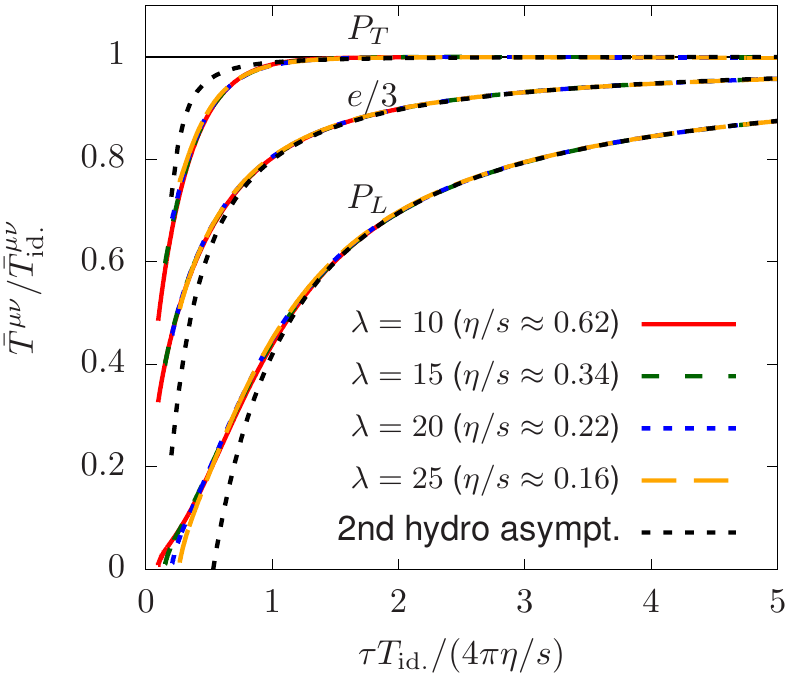}
\end{minipage}
\begin{minipage}{2.5in}
\includegraphics[width=2.5in]{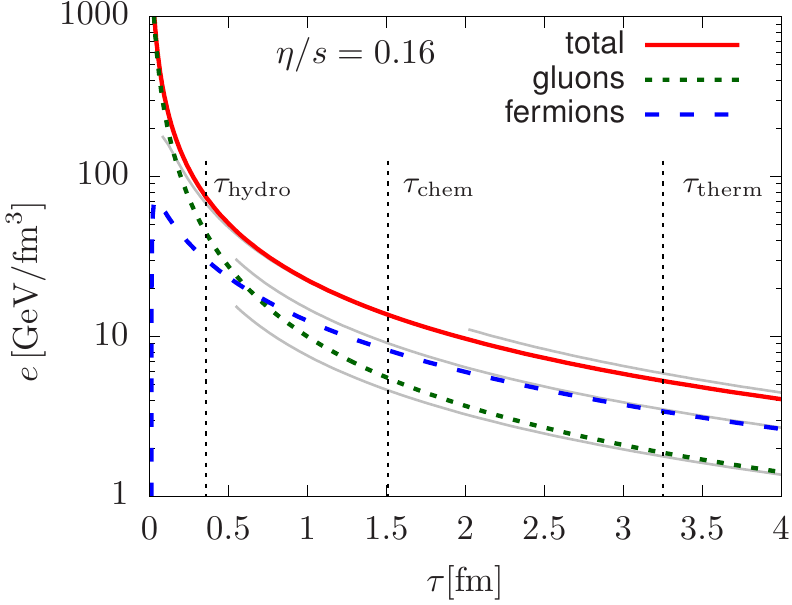}
\end{minipage}
\caption{\label{fig:QuarkGluon} (left) Non-equilibrium evolution of the
different components of the average energy-momentum tensor
$T^{\mu\nu}={\rm diag}(e,P_T,P_T,P_L)$ compared to viscous fluid dynamics
\cite{Kurkela:2018vqr,Kurkela:2018wud}. (right) Evolution of the overall energy
density $e$ and the energy density carried by quarks and gluons $e_{g/q}$
\cite{Kurkela:2018xxd}.}
\end{figure}

Neglecting potential problems related to plasma instabilities, the
non-equilibrium evolution of macroscopic quantities such as 
$T^{\mu\nu}$ can be calculated based on numerical simulations of the
effective kinetic theory.
Numerical simulation based on QCD kinetic theory were pioneered in \cite{Xu:2004mz,El:2007vg}; the first complete leading order study for a homogeneous purely
gluonic plasma was performed in \cite{Kurkela:2015qoa} and subsequently extended to inhomogeneous plasmas \cite{Keegan:2016cpi,Kurkela:2018vqr,Kurkela:2018wud} as well as
homogeneous plasmas of quarks and gluons \cite{Kurkela:2018xxd,Kurkela:2018oqw}. 
Kinetic theory simulations shown in Fig.~\ref{fig:QuarkGluon}(a) indicate that for
realistic coupling strength $\alpha_s \gtrsim 0.1$, the evolution of the energy-momentum tensor towards equilibrium is to a good approximation controlled by
a single time scale $\tau^{\rm eq}_{R}$, corresponding to the \emph{equilibrium
relaxation rate}
\begin{eqnarray}
\tau_{R}^{\rm eq}(\tau)= \frac{4 \pi \eta/s}{T_{\rm Id}(\tau)}\;,
\end{eqnarray}
where $\eta/s \propto \lambda^2 $ is the shear-viscosity to entropy density ratio, and $T_{\rm Id}(\tau)=
\propto \tau^{-1/3}$ denotes the temperature of the late-time equilibrium
system. Even though an extrapolation to sizeable coupling strength is required
to make contact with heavy-ion phenomenology, the dependence on 
$\alphas$ is surprisingly weak once $\tau$ is measured in units of
$\tau^{\rm eq}_{R}$. When comparing the results
for the non-equilibrium evolution of the energy-momentum tensor $T^{\mu\nu}$ in
kinetic theory with the asymptotic behavior in viscous hydrodynamics, one
concludes that a fluid dynamic description becomes applicable on time scales
$\tauhydro \approx \tau^{\rm eq}_{R}(\tau)$. For phenomenological
purposes the coupling constant $\lambda$  
can be traded for $\eta/s\propto \lambda^2$ 
yielding the following estimate
\begin{equation}
\label{eq:hydrotime}
\tauhydro\approx 1.1\,{\rm fm} \, \left( \frac{4\pi(\eta/s)}{2} \right)^{{3}/{2}}  \left( \frac{ \langle \tau s\rangle }{ 4.1 \, {\rm  GeV}^2 } \right)^{-1/2} \left( \frac{\nu_\text{eff}}{40} \right)^{1/2},
\end{equation}
where $\langle \tau s\rangle$  denotes the entropy density per unity rapidity.
$\tau s$ is directly related to the charged particle multiplicity
$dN_{\rm ch}/d\eta$, and thereby constrained to be approximately $\approx 4.1\,{\rm GeV}^{2}$ for
central $\textrm{Pb+Pb}$ collisions at LHC energies \cite{Keegan:2016cpi}. Since the
discussion so far ignores the effects of spatial gradients, both in transverse
space and longitudinal rapidity, the estimate in (\ref{eq:hydrotime}) should be
understood as a lower bound. 

Interestingly one finds that viscous hydrodynamics starts to describe the
evolution of the energy-momentum tensor in a regime where both the Knudsen
number ${\rm Kn}_{\theta} \approx \tau_{R}^{\rm eq}/\tau$  and the proximity
to equilibrium as measured by  $1-P_{L}/P_T$ are of order unity, 
indicating that the system is still significantly out-of-equilibrium.
Even though this behavior appears to be quite surprising, it is by no means
unique to a weakly coupled non-equilibrium description, and similar observations
have been reported much earlier in the context of strongly coupled gauge
theories \cite{Heller:2011ju}. It has become common to
distinguish  the time when hydrodynamics becomes applicable $\tauhydro$  
(the so called ``hydrodynamization'' time)
from the time $\tau_{\rm eq}$ when the pressure anisotropy is small. Due to
the rapid longitudinal expansion, the actual approach towards 
local pressure isotropy
occurs only on much larger time scales $\tau_{\rm eq} \gg \tauhydro$.
Hence the great success of hydrodynamic descriptions of the QGP does not appear
to derive from the fact that the system is particularly close to equilibrium
throughout most of its space-time evolution, but is rather due to fact that the
range of applicability of viscous relativistic fluid appears to be larger than
originally anticipated. Notably, these observations have triggered a large
number of theoretical studies to further investigate and possibly extend the
range of applicability of viscous fluid dynamics~\cite{Florkowski:2017olj,Romatschke:2017vte,Strickland:2017kux}. However, a detailed discussion of these topics is beyond the scope of this review.

\subsection{Quark production and chemical equilibration}
So far most theoretical studies of the early non-equilibrium dynamics have
focused on the kinetic equilibration of gluons, while neglecting dynamical
fermions in the analysis. However, on a conceptual level it is equally
important to understand the transition from an initial state, which is believed
to be highly gluon dominated, towards chemical equilibrium where a significant
fraction of the energy density is carried by quark degrees of freedom. We note
from a phenomenological point of view the chemical composition of the plasma at
early times, may have also have interesting consequences, e.g. relating to the
questions concerning the chemical equilibration of strange quarks and heavy
flavors or the electro-magnetic response of the QGP at very early times after
the collision. Even though a complete picture of chemical equilibration along
the lines of our discussion in Sec.~\ref{basics_underoccupied} is yet to be established, interesting
first results have been reported in the literature. 
We briefly discuss these results below.

Classical-statistical simulations of quark production at very early times have
been pioneered in \cite{Gelis:2005pb} demonstrating that at realistic coupling
strength a significant number of quark anti-quark pairs can be produced in the
initial (semi-) hard scattering and in the presence of the strong color fields
at very early times. Subsequent studies have refined the lattice approach
\cite{Gelis:2015eua,Mueller:2016ven} and further elaborated on quark production
in over-occupied systems \cite{Tanji:2017xiw}. However, as
classical-statistical simulations involving dynamical fermions are
significantly more complicated, studies are yet to reach the same level of
sophistication of analogous pure gauge theory simulations.

Quark production during the final radiative break-up stage of the bottom up
scenario, has been investigated in the context of
jet quenching~\cite{Mehtar-Tani:2018zba}, where it was pointed out that the turbulent nature of the inverse energy cascade ultimately determines the quark/gluon ratio from a local balance of the $g\to q \bar{q}$ and $q\to q g$ processes. However,
within the inertial range of  the cascade $T_{\rm soft} \ll p \ll p_{\rm
split}$ the fraction of energy carried by quarks and anti-quarks 
$e_{\rm q}/e_{\rm g}  \simeq 0.07 \times 2N_f$ (for three colors) is small compared to
the  equilibrium ratio 
$e_{\rm q}/e_{\rm g}  \simeq 0.3 \times 2N_f$,
indicating
that elastic processes, which are operative at the scales of the soft thermal
medium also play a pivotal role in the chemical equilibration process.

The first numerical study implementing all relevant leading
order processes of bottom up was performed in \cite{Kurkela:2018xxd,Kurkela:2018oqw},
indicating that as shown in Fig.~\ref{fig:QuarkGluon} the approach to viscous
fluid dynamic behavior (discussed in \Sect{sec:hydrodynamization}) occurs \emph{before} chemical
equilibration of the QGP. 
A complete leading order  analysis of the chemical equilibration mechanism (along the lines of \Sect{basics_underoccupied}) has not yet been given, and should explain these first numerical results and provide guidance to phenomenology.

Notably, the inclusion of dynamical quarks also represent an important step
towards calculations of pre-equilibrium photon and dilepton production, and in
addressing questions related to the chemical/kinetic equilibration of heavy
flavors. While first progress in this direction has been reported in
\cite{Berges:2017eom} by analyzing a subset of leading order processes, a
complete leading order study has not been performed to date.

\subsection{Small scale fluctuations and pre-flow}
\label{sec:preflow}

So far we have discussed the microscopic dynamics of the local equilibration
process, neglecting the effects of spatial gradients on small scales $\sim c
\tau_{\rm hydro}$. However, as discussed in \Sect{Bottomup} the inclusion of small
scale fluctuations $\sim R_{p}$ is a necessary ingredient for a realistic
event-by-event description, since such gradients will lead to the
development of ``pre-flow'', a pre-cursor to the late stage hydrodynamic
flow which starts to build up already during the
pre-equilibrium phase. The kinetic theory should evolve these fluctuations 
and smoothly asymptote to hydrodynamics at late times $\tau \sim \tau_{\rm hydro}$.

\begin{figure}
   \begin{center}
\begin{minipage}{3.0in}
\includegraphics[width=2.5in]{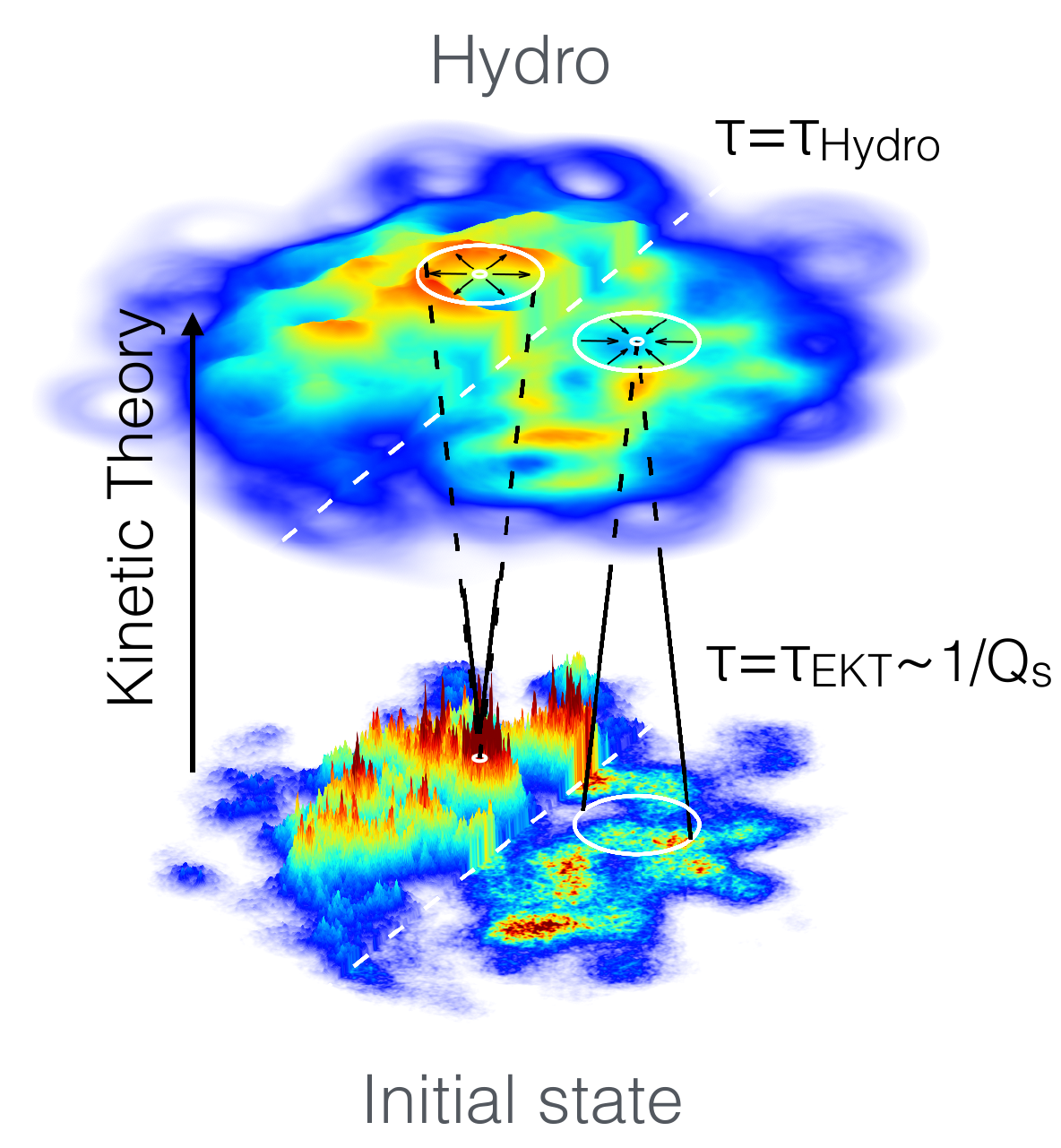}
\end{minipage}
\caption{\label{fig:KompostFIG} Schematic of the transverse energy density profiles at very early times $(\tau_{0}=\tau_{\scriptscriptstyle {\rm EKT}} \approx 0.1\,{\rm fm}/c)$ and after the first ${\rm fm}/c$ of pre-equilibrium evolution done with kinetic theory $(\tau=\tauhydro)$. At a time $\tau_{\rm hydro}$ the constitutive equations are approximately satisfied (see \Fig{fig:QuarkGluonKompost}). }
\end{center}
\end{figure}

A recent extension of
the bottom up scenario accounts for  small scale fluctuations
by explicitly including spatially inhomogeneous fluctuations of the phase space
density into the kinetic description~\cite{Kurkela:2018vqr,Kurkela:2018wud,Keegan:2016cpi}.
By choosing a representative form for the
phase-space distribution to model the initial fluctuations  of the
stress tensor $\delta T^{\mu\nu}(\tau_0,\x_0)$ around a local average  $\bar{T}^{\mu\nu}_{\bf x}(\tau_0)$ at
a point ${\bf x}$, 
the pre-equilibrium evolution of the energy-momentum tensor can then be calculated
as
\begin{eqnarray}
\label{eq:linrespo}
T^{\mu\nu}(\tau,{\bf x}) \simeq \underbrace{\bar{T}^{\mu\nu}_{\bf x}(\tau)}_{\substack{\text{non-eq. evolution of} \\ \text{(local) avg. background}}} + 
\underbrace{\int_{\odot} d^2{\bf x}_{0}~G^{\mu\nu}_{\alpha\beta}(\tau,\tau_0,{\bf x},{\bf x}_0)~\delta T^{\alpha\beta}(\tau_0,{\bf x}_0) }_{\text{non-eq. evolution of local fluctuations of the stress tensor}}\;,
\end{eqnarray}
which is shown schematically in \Fig{fig:KompostFIG}. 
Here $\bar{T}^{\mu\nu}_{\bf x}(\tau)$  describes the
pre-equilibrium evolution of the average energy-momentum tensor and is described by
Fig.~\ref{fig:QuarkGluon}, while  $G^{\mu\nu}_{\alpha\beta}$ describes the  linear response to initial fluctuations in
the thermalizing plasma~\cite{Kurkela:2018vqr}.
Since causality restricts the contributions to the fluctuations at ${\bf x}$, one only needs to integrate the response over the causal circle $\odot$ indicated by the circle at $\tau_0=\tau_{\scriptscriptstyle {\rm EKT}}$ in \Fig{fig:KompostFIG}. 
The relevant response functions $\bar{T}^{\mu\nu}(\tau)$
and $G^{\mu\nu}_{\alpha\beta}$, can be calculated once and for all in kinetic
theory, and packaged into a useful ``pre-flow'' computer code which encapsulates
the thermalization process~\cite{Kurkela:2018vqr}.

\begin{figure}
\begin{minipage}{2.5in}
\includegraphics[width=2.5in]{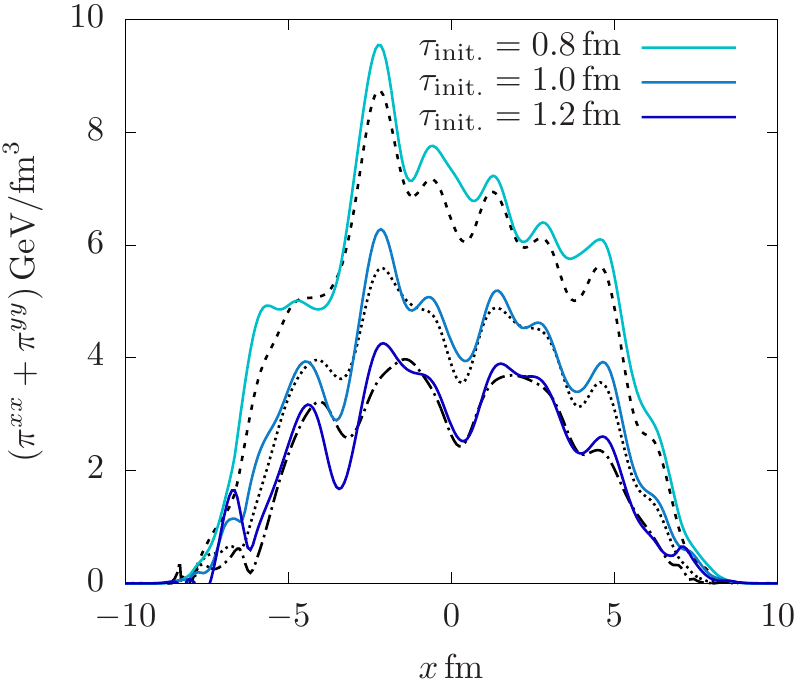}
\end{minipage}
\begin{minipage}{2.5in}
\includegraphics[width=2.5in]{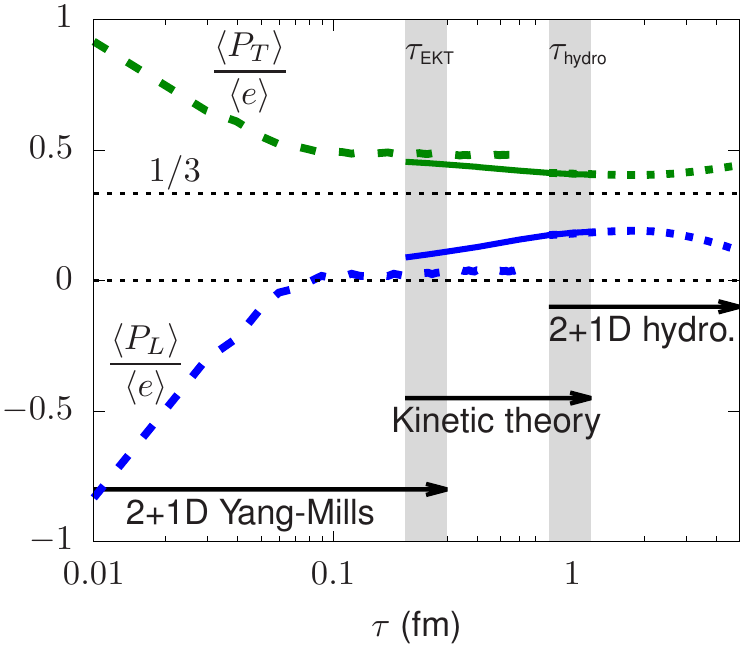}
\end{minipage}
\caption{\label{fig:QuarkGluonKompost} 
   (left) Spatial profiles of the non-equilibrium shear-stress tensor $(\Pi_{xx}+\Pi_{yy})$ (the solid lines) compared to the Navier-Stokes hydrodynamics limit (the dashed lines) after an evolution of $\tau_{\rm init.}$ in the kinetic theory (see \Eq{eq:linrespo}).
(right) 
Proof of principle calculation combining different theoretical descriptions to calculate the evolution of energy density $e$,$P_T$,$P_L$ in a single ${\rm Pb}{+}{\rm Pb}$ event~\cite{Kurkela:2018vqr,Kurkela:2018wud}.  
}
\end{figure}

The linear response formalism of \Eq{eq:linrespo}
can be seen as a
systematic extension of earlier studies \cite{Vredevoogd:2008id}, recognizing
universal patterns in the pre-equilibrium evolution of the long wave-length
components of the energy-momentum tensor. Short wave-length fluctuations $\ll c
\tau_{\rm Hydro}$ are efficiently damped during the pre-equilibrium phase,
leading to an effective coarse graining of the spatial profile of the energy-momentum tensor shown schematically in \Fig{fig:KompostFIG}. 
Then viscous corrections to the energy-momentum tensor are reasonably well approximated by
the Navier-Stokes constitutive relations  at the time $\tau_{\rm init.}$ when hydrodynamics
is initialized. This is shown in \Fig{fig:QuarkGluonKompost}\,(left),  which uses \Eq{eq:linrespo}
to determine the stress at a time $\tau_{\rm init.}$.
Long wave-length fluctuations of the initial energy density determine the pre-flow which develops during thermalization
process, and  can be reasonably approximated as 
\begin{eqnarray}
	T^{\tau i}(\tau,{\bf x}) \approx -\frac{(\tau-\tau_0)}{2}  \left(\frac{\bar{T}_{\bf x}^{\tau\tau}(\tau)}{\bar{T}_{\bf x}^{\tau\tau}(\tau_0)} \right) \partial^i\bar{T}^{\tau\tau}(\tau_0,\x)\, .
\end{eqnarray}
Nevertheless, the results of \cite{Kurkela:2018vqr,Kurkela:2018wud} also demonstrate that a genuine
non-equilibrium description is necessary account for the entropy production during the
pre-equilibrium phase. Since the subsequent hydrodynamic expansion
approximately conserves the overall entropy, this factor two to three increase
in entropy during the pre-equilibrium phase is important in relating properties
of the initial state to experimentally observed charged particle
multiplicities.

Finally, by combining the classical-statistical field simulations 
at early times, the kinetic simulations at intermediate
times, and the hydrodynamics simulations at late times, a consistent space-time description of
the energy-momentum tensor can  be obtained on an
event-by-event basis. This is illustrated in Fig.~\ref{fig:QuarkGluonKompost} (right) which shows
the evolution of longitudinal and transverse pressures in a single
${\rm Pb}{+}{\rm Pb}$ event. In this simulation the first stage up to $\tau_{\rm EKT}$ is treated in the classical IP-Glasma model (see \Sect{Bottomup}); the second stage up to $\tau_{\rm hydro}$ is treated with QCD kinetics following
the outlines of bottom-up;
and the final phase is treated with hydrodynamics. The different theoretical descriptions overlap providing
a complete picture of the event.

\section{Outlook and small systems}
\label{sec:epilogue}

We have reviewed the weak coupling description of the thermalization process of the QGP during the first fm/c of high-energy heavy-ion collisions, by dividing the out-of-equilibrium dynamics of non-abelian gauge theories into two broad classes -- an over-occupied limit discussed in \Sect{basics_overoccupied}, and an under-occupied limit discussed in \Sect{basics_underoccupied}. Strikingly, the thermalization process in each of these limits exhibits generic scaling features which one would like to observe experimentally.

Indeed, much of the current interest in the equilibration process is driven by
exciting new data on the small systems created in proton-proton ($p{+}p$) and
proton-nucleus ($p{+}A$) collisions, which show evidence for a transition
towards a hydrodynamic regime in nucleus-nucleus $A{+}A$ collisions. A more complete review of the experimental data is given in the literature~\cite{Loizides:2016tew,Dusling:2015gta}. In the
larger $A+A$ system, the approach to hydrodynamics has been largely understood
and quantified within the bottom-up scenario (see for example \Eq{eq:hydrotime}
of \Sect{sec:pheno}), and the physics of the pre-equilibrium stage has been
packaged into a useful ``pre-flow'' computer code that can be used to simulate
heavy ion events (see \Sect{sec:preflow}). However, as the system size gets
smaller, additional scales, such as the transverse radius $R$, play an
increasingly
important role and truncate the thermalization process. Nevertheless, one can use the bottom-up framework to estimate  
when hydrodynamics becomes applicable as a function of the multiplicity produced in
the collision~\cite{Kurkela:2018vqr}. Substituting $\tau s = dS/dy /\pi R^2$ in \Eq{eq:hydrotime} we find
\st
\frac{\tau_{\rm hydro}}{\rm R} \simeq \left(\frac{dN_{\rm ch}/dy}{63} \right)^{-1/2} \left(\frac{4\pi \eta/s}{2} \right)^{3/2}
\left( \frac{S/N_{\rm ch} }{7} \right)^{-1/2} \left( \frac{\nu_{\rm eff} }{40} \right)^{1/2} \, .
\stp
Since we expect the bottom up analysis will be strongly modified for $\tau_{\rm hydro}/R \sim 1$,  a charged particle multiplicity 
of order $dN_{\rm ch}/dy \sim 70$ should demarcate the transition to a fully equilibrated regime.  So far a detailed understanding,
both parametrically and numerically of the transition regime has not been given, though important first steps
have been taken~\cite{Greif:2017bnr,Borghini:2018xum,Kurkela:2018ygx}.  In small systems there are by now many experimental tools,
(such as e.g. the hadron chemistry~\cite{Kurkela:2018xxd} or the system size dependence of the harmonic flow~\cite{Yan:2013laa,Mace:2018yvl})
which can be used to clarify the kinetics of high energy QCD and to guide
theory. Further as emphasized in \Sect{sect_cascade}, studies of the energy
loss of jets, both in small and large systems, can inform the study of
thermalization of QCD plasmas.  We therefore anticipate that, through  a
combination of phenomenology, formal theory, experiment, and simulation, the
community will analyze the transition from cold QCD to the hot QGP in detail,
and, more generally, clarify the out-of-equilibrium behavior of non-abelian gauge
theories.

\section*{DISCLOSURE STATEMENT}
The authors are not aware of any affiliations, memberships, funding, or financial holdings that might be perceived as affecting the objectivity of this review. 

\section*{ACKNOWLEDGMENTS}
We gratefully acknowledge helpful discussions with Peter Arnold, Juergen
Berges, Aleksi Kurkela, Aleksas Mazeliauskas, Jean-Francois Paquet, and Raju
Venugopalan. This work was supported in part by the U.S. Department of Energy,
Office of Science, Office of Nuclear Physics  under Award Numbers
DE\nobreakdash-FG02\nobreakdash-88ER40388 and by the Deutsche
Forschungsgemeinschaft (DFG, German Research Foundation) – Project number
315477589 – TRR 211.

\bibliographystyle{ar-style5}
\bibliography{allrefs}

\end{document}